# SHORTFALL DEVIATION RISK: AN ALTERNATIVE FOR RISK MEASUREMENT

**Marcelo Brutti Righi**
**Paulo Sergio Ceretta**

## Abstract

We present the Shortfall Deviation Risk (SDR), a risk measure that represents the expected loss that occurs with certain probability penalized by the dispersion of results that are worse than such an expectation. SDR combines Expected Shortfall (ES) and Shortfall Deviation (SD), which we also introduce, contemplating two fundamental pillars of the risk concept – the probability of adverse events and the variability of an expectation – and considers extreme results. We demonstrate that SD is a generalized deviation measure, whereas SDR is a coherent risk measure. We achieve the dual representation of SDR, and we discuss issues such as its representation by a weighted ES, acceptance sets, convexity, continuity and the relationship with stochastic dominance. Illustrations with real and simulated data allow us to conclude that SDR offers greater protection in risk measurement compared with VaR and ES, especially in times of significant turbulence in riskier scenarios.

**Keywords:** Shortfall deviation risk, risk management, risk measures, coherent risk measures, generalized deviation measures.

## 1. Introduction

Risk is an important financial concept due to the influence it has on other concepts. During times of financial turbulence (e.g., crises and collapses in the financial system) the focus on risk management increases. A fundamental aspect of appropriate risk management is measurement, especially forecasting risk measures. Risk overestimation can cause an agent to retain capital that can be applied to profitable investments, whereas underestimating risk can cause very large shortfalls, which can be alarming to an agent.

Following a notable study by Markowitz (1952), the risk of a financial position has begun to be addressed more scientifically. The use of variability measures, such as variance, semi-deviation and standard deviation, have become common for representing risk. With the development and integration of financial markets and the occurrence of critical events, the need has emerged for another type of measurement based on the larger and least likely losses, which are known as tail risks. The commercial product RiskMetrics, which was introduced by J.P. Morgan in the early 1990s, employs a risk measure that is based on the distribution quantile of results, known as Value at Risk (VaR). VaR represents a loss that is only surpassed given a significance level for a certain period. VaR has become the standard measure of financial risk since it was sanctioned by the Basel Committee, an entity that serves as risk management practices counsel for financial institutions. Duffie and Pan (1997) and Jorion (2007) have examined VaR in their studies.

Despite the extensive practical use of risk measures, few studies have defined the characteristics of a desirable risk measure. Literature has emerged that discusses, proposes and criticizes the theoretical properties of a risk measure. To fill this gap, a class of coherent risk measures was developed and introduced by Artzner et al. (1999), who present axioms for a risk measure for use in practical matters. Thus, theoretical discussions on risk measures began to gain attention in the literature, with less emphasis being devoted to analyses that were only conducted from an empirical perspective. Other risk measure classes have been developed, including convex measures, which were simultaneously presented by Föllmer and Schied (2002) and Frittelli and Gianin (2002), spectral measures, which were proposed by Acerbi (2002), and generalized deviation measures, which were introduced by Rockafellar et al. (2006).

Based on the coherence axioms and other risk measure classes, the indiscriminate use of VaR began to receive considerable criticism because it is not a convex measure, which implies that the risk of a diversified position can be greater than the sum of individual risks. In addition, VaR completely disregards the potential of losses beyond the quantile of interest, which can be very dangerous. Some studies address the lack of coherence of VaR and present alternatives that satisfy the axioms. Thus, the expected value of losses that exceeds VaR was defended as a potential risk measure. Different authors have proposed similar concepts using different names to fill the identified gap. Acerbi and Tasche (2002a) present the Expected Shortfall (ES), Rockafellar and Uryasev (2002) and Pflug (2000) introduce the Conditional Value at Risk (CVaR), Artzner et al. (1999) argue for the Tail Conditional Expectation (TCE), which is also referred to as the Tail Value at Risk (TVaR), and the Worst Conditional Expectation (WCE), Longin (2001) presents the Beyond Value at Risk (BVaR), Föllmer and Schied (2011) refer to it as the Average Value at Risk (AVaR).

These studies indicate the advantages of the proposed measure compared to VaR. Although these measures have extremely similar definitions, ES is the more relevant coherent risk measure in finance. As demonstrated by Acerbi and Tasche (2002a), these definitions produce the same results when applied to data with continuous distributions. These authors show that the advantage of the ES definition is its coherence regardless of the underlying distribution and its effectiveness in estimation even in cases where VaR estimators can fail. Acerbi and Tasche (2002b) discuss the properties of ES coherence and compare several representations of the measure, which are more suitable for certain proposals. In the search for an adequate alternative to VaR, Tasche (2002) notes that ES has been characterized as the most relevant risk measure.

Despite the advantages of ES, this measure is less frequently utilized than VaR because forecasting ES is challenging due to its complex mathematical definition. Studies have been conducted to compare VaR and ES, in addition to other measures, with divergent results regarding the advantages in their application. Although VaR is flawed in many cases, implementation of ES is difficult or does not address the entire definition of risk, as suggested by Yamai and Yoshiba (2005), Dowd and Blake (2006) and Guégan and Tarrant (2012). Other studies indicate that VaR is not so bad and the use of ES can produce poor results, as noted by Alexander and Baptista (2004), Dhaene et al. (2008), Wylie et al. (2010), Bamberg and Neuhierl (2010) and Daníelsson et al. (2013).

Due to the nonexistence of a consensus regarding an appropriate measure, there is space for other risk measures. Coherent measures, such as the Weighted Value at Risk (WVaR) proposed by Cherny (2006), which is a weighted version of measures such as ES and CVaR, are discussed in the literature. Föllmer and Knispel (2011) present a coherent version of the entropic measure, which is based on an exponential utility function. Another example is the Entropic Value at Risk (EVaR) measure proposed by Ahmadi-Javid (2012), which corresponds to the most restrictive upper limit obtained by an inequality between VaR and CVaR. Jadhav et al. (2013) present the Modified Expected Shortfall (MES), which represents the expected value of losses that are situated between the VaR and the negative of another quantile of interest and is always smaller than ES. Fischer (2003) considers unilateral measures, which combine the mean and semi-deviation at higher moments, whereas Chen and Wang (2008) consider bilateral measures that also incorporate the semi-deviation of gains. Krokhmal (2007) also considers higher moments to obtain coherent risk measures as solutions for optimization problems that consider the dispersion of losses. Fölmer and Schied (2002) present the Shortfall

Risk (SR), which is the expected value of a convex and increasing loss function. Belles-Sampera et al. (2014) present the glue value at risk (GLUEVaR), which is a combination of TVaR for two different quantiles with VaR of one of the two quantiles. Bellini et al. (2014) advocate the use of expectiles because they comprise the only generic representation of quantile functions that satisfy the coherence properties.

Given the need for new measures, even approaches that do not guarantee coherence or escape from traditional focus have been proposed. Jarrow (2002) presents a measure based on put options premium. Bertsimas et al. (2004) suggests the concept of the difference between expected loss, and ES. Chen and Yang (2011) introduce the Weighted Expected Shortfall (WES), which is an ES version that assigns different nonlinear weights to losses that exceed the VaR. Belzunce et al. (2012) use the ratio between ES and VaR, which is referred to as the Proportional Expected Shortfall (PES), to obtain a universal measure for risks of different natures. To extend to a multivariate dimension, Cossette et al. (2013) and Cousin and Di Bernardino (2013) present the lower orthant VaR and the upper orthant VaR, which represent the maximum VaR and the minimum VaR, respectively, of a set of assets. Similarly, Cousin and Di Bernardino (2014) extend the upper and lower orthant VaR concept to the TCE case. Prékopa (2012) defines the multivariate VaR as a set of quantiles of a multivariate probability distribution. Lee and Prékopa (2013) develop the theory and methodology of the multivariate VaR and CVaR based on adaptations of multivariate quantiles. Hamel et al. (2013) present the multivariate AVaR defined in sets, rather than scalar or even a vector.

Despite the proposition of other risk measures for financial markets, these measures did not achieve the same amount of success as VaR or ES due to their complexity or their close relation to ES. The expected value of losses has become the primary focus. However, the variability concept, which is one of the risk concept pillars, is disregarded in this definition of risk measure. The central focus of this study is to propose a risk measure that includes the dispersion degree of an extreme loss jointly to its expected value in measuring risk. Because two financial positions with the same expected return may exhibit different variability when all available data are considered, discrepancies may also result if only the extreme values are considered.

In this study, we consider dispersion, which is measured by the semi-deviation of results that represent losses greater than ES. This deviation is referred to as Shortfall Deviation (SD) in this study. Using the concepts of ES and SD, this study aims to introduce a new risk measure, the Shortfall Deviation Risk (SDR), which can be defined as the expected loss, when it exceeds VaR, penalized by the dispersion of results that represent losses greater than this expectation. Although this characteristic is disregarded by ES and related measures, it is encompassed by SDR. In addition to combining the two fundamental risk concepts, the probability of poor results (ES) and the variability of an expected result (SD), into a single measure, SDR considers the tails, which represent turbulence moments, where proper risk management is most needed. Thus, the SDR can be described as a more complete measure in that sense. The SDR exceeds ES because it yields higher values due to penalty by dispersion and it may serve as a more solid protection barrier. Based on this perspective, we discuss in detail the SDR definition and prove its theoretical properties. SD is a generalized deviation measure, as proposed by Rockafellar et al. (2006), whereas SDR is a coherent risk measure in the sense of Artzner et al. (1999). In addition to its concrete practical definition, SDR possesses solid theoretical properties that ensure its use without violating axiomatic assumptions. An illustration using simulations and real data present the proposed concepts in a more practical way.

We contribute to academic literature and financial industry because we propose a new risk measure, the SDR. The tail dispersion concept is not new because variance, and consequently, standard deviation of a truncated distribution, is a well-established concept. However, studies such as Wu and Xiao (2002), Bali et al. (2009), Righi and Ceretta (2013) and Righi and Ceretta (2015) do not discuss any theoretical properties that support its use. Valdez (2005) and Furman and Landsman (2006a) discuss theoretical properties of the tail variance but they do not fit it to risk measure classes. These studies consider the variance that is truncated by VaR, equally penalizing losses that are higher and lower than the ES. Because the idea is to penalize the risk measured by ES, it is reasonable to consider the dispersion of results that represent losses greater than ES. Wang (1998) presents a tail deviation that represents the difference between distorted and undistorted expectations, demonstrating its properties. Sordo (2009) expands the concept to additional tail dispersion forms in addition to the standard deviation. Our study advances this stream because we present in a more complete fashion the SD component characteristics as a generalized deviation measure.

Fischer (2003) and Chen and Wang (2008) consider combining mean and semi-deviations to different powers to form a coherent risk measure. However, SDR is defined for the tails, unlike the measures proposed by these authors. Krokhmal (2007) extends the ES concept, obtained as the solution to an optimization problem, for cases with higher moments to demonstrate that they are coherent risk measures and have a relationship with generalized deviation measures. However, measurements obtained by the method proposed by this author do not have an explicit financial meaning, such as the SDR, which enables the value obtained to be intuitively defined. Furman and Landsman (2006a, 2006b) propose a measure that is similar to SDR, which weighs the mean and standard deviation in the truncated tail by VaR, and discuss some theoretical properties. The SDR, however, considers higher losses than ES for penalty, which results in better properties, such as positions with higher losses to pose higher risk, unlike the measure presented by the authors. In addition, the weighting scheme of SDR is different, which ensures greater penalty for deviation in more extreme quantiles. Thus, SDR is superior to that measure because it has a large number of theoretical properties and is classified as a coherent risk measure.

The remainder of this study is structured as follows. Section 2 presents notation, definitions and preliminary results. Section 3 presents the main results. Section 4 describes illustrations based on simulated and real data. Section 5 presents the conclusions of the study and advocates for the application of the SDR for different fields in finance.

## 2. Theoretical Background

Unless otherwise stated, the content is based on the following notation. Consider a single-period market, with a current date 0 and a future date $T$. No transaction is possible between 0 and $T$. Consider the random result $X$ of any asset or portfolio that is defined in an atomless probability space $(\Omega, \mathcal{F}, \mathbb{P})$. Thus, $E_{\mathbb{P}}[X]$ is the expected value of $X$ under $\mathbb{P}$. In addition, $\mathcal{P} = \{\mathbb{Q} | \mathbb{Q} \ll \mathbb{P}\}$, where $\ll$ stands for absolute continuity, is a nonempty set, because $\mathbb{P} \in \mathcal{P}$, which represents the probability measures $\mathbb{Q}$ defined in $\Omega$ that are absolutely continuous in relation to $\mathbb{P}$. $\frac{d\mathbb{Q}}{d\mathbb{P}}$ is the density of $\mathbb{Q}$ relative to $\mathbb{P}$, which is known as the Radon-Nikodym derivative. All equalities and inequalities are considered to be almost surely (a.s.) in $\mathbb{P}$. $F_X$ is the probability function of $X$ and its inverse $F_X^{-1}$. Because $(\Omega, \mathcal{F}, \mathbb{P})$ are atomless, $F_X$ can be assumed to be continuous, and this assumption is made throughout the study. Let $L^p = L^p(\Omega, \mathcal{F}, \mathbb{P})$ be the space of random variables of which $X$ is an element, with $1 \leq p \leq \infty$, as defined by the norm $\|X\|_p = (E_{\mathbb{P}}[|X|^p])^{\frac{1}{p}}$ with finite $p$ and $\|X\|_p = \inf\{k : |X| \leq k\}$ for infinite $p$. $X \in L^p$ indicates that $\|X\|_p < \infty$, implying that the absolute value of $X$ to the $p$ power is limited and integrable. Furthermore, $(X)^- = \max(-X, 0)$. In this context, measuring risk is equivalent to establishing the function $\rho : L^p \to \mathbb{R}$, in other words, summarizing the risk of position $X$ into one number.

Here, we present definitions and results present in the literature, which facilitate the development of the theoretical framework of the proposed measure. Initially, we first define, in addition to SD and SDR, VaR and ES concepts, which are critical to understanding the proposed measure.

**Definition 1**. *Let $X \in L^p$. Given a significance level $0 \leq \alpha \leq 1$:*

$$VaR^\alpha(X) = - \inf \{x : F_X(x) \geq \alpha\} = -F_X^{-1}(\alpha) = -q_\alpha(X). \tag{1}$$

$$ES^\alpha(X) = -E_\mathbb{P}[X | X \leq q_\alpha(X)] = -e_\alpha(X). \tag{2}$$

$$SD^\alpha(X) = \left(E_\mathbb{P}\left[\left|\left(X - e_\alpha(X)\right)^-\right|^p\right]\right)^{\frac{1}{p}} = \left\|\left(X - e_\alpha(X)\right)^-\right\|_p. \tag{3}$$

$$SDR^\alpha(X) = ES^\alpha(X) + (1 - \alpha)^\beta SD^\alpha(X), \beta \geq 0. \tag{4}$$

**Remark 1**. VaR is the quantile $q_\alpha$ of $X$, which is adjusted by the negative sign and represents a loss between 0 and $T$ that it is only exceeded with probability $\alpha$. VaR does not consider information after the quantile of interest, only the point itself. ES overcomes this difficulty because it represents the expectation of $X$, adjusted by the negative sign, conditioned to $X$ represents a higher loss than VaR, i.e., an extreme loss. We propose that the dispersion truncated by ES is considered as a penalty term. This measure is the SD, which is the semi-deviation in relation to the ES. Distinct values of $p$ can incorporate higher moments of $X$. Based on these definitions, we develop a risk measure that adjusts the risk of extreme losses through its dispersion. Two positions can present the same tail expected loss but different dispersions. While one position has a certain expected loss, another position may present dispersion in such a manner that excessively higher losses can occur. Based on this reasoning, the SDR rises.

**Remark 2**. SDR simultaneously encompasses two risk definition pillars because it considers the possibility of extreme bad results and the uncertainty relative to an expected value. The term $(1 - \alpha)^\beta$ represents how much dispersion should be included as an ES penalty, which may serve as protection. Lower values of $\beta$ generate higher penalties, with the minimum case of $(1 - \alpha)^\beta = 0$ with $\beta = \infty$, in which the original value of ES is recovered, and the maximum case of $(1 - \alpha)^\beta = 1$ with $\beta = 0$, in which all SD is incorporated. The choice of values for $\beta$ enables the incorporation of subjective issues, such as the degree of risk aversion of the agent. Moreover, SDR yields higher values than ES and VaR and lower values than the maximum loss $\sup -X$; it is also nonincreasing in $\alpha$ because the risk measure has greater values in more extreme quantiles.

The risk measure class in which SDR fits consists of the coherent risk measures proposed by Artzner et al. (1999). Thus, the aim is to prove axioms of this class for SDR in addition to other properties and characteristics such as acceptance sets and dual representation. We now define coherent risk measures.

**Definition 2**. *A function $\rho : L^p \to \mathbb{R}$ is a coherent risk measure if it fulfills the following axioms:*

*Translation Invariance: $\rho(X + C) = \rho(X) - C, \forall X \in L^p, C \in \mathbb{R}$.*
*Subadditivity: $\rho(X + Y) \leq \rho(X) + \rho(Y), \forall X, Y \in L^p$.*
*Monotonicity: if $X \leq Y$, then $\rho(X) \geq \rho(Y), \forall X, Y \in L^p$.*
*Positive Homogeneity: $\rho(\lambda X) = \lambda \rho(X), \forall X \in L^p, \lambda \geq 0$.*

*Additionally, a coherent risk measure can respect the following axioms:*

> *Relevance: if $X \leq 0$ and $X \neq 0$, then $\rho(X) > 0, \forall X \in L^p$.*
> *Strictness: $\rho(X) \geq -E_{\mathbb{P}}[X], \forall X \in L^p$.*
> *Law Invariance: if $F_X = F_Y$, then $\rho(X) = \rho(Y), \forall X, Y \in L^p$.*

**Remark 3**. The first axiom ensures that if a certain gain is added to a position, its risk should decrease in the same amount. The second axiom, which is based on the principle of diversification, implies that the risk of a combined position is less than the sum of the individual risks. The third axiom requires that if first position always generates worse results than the second position, the risk of the first position shall always be greater than the risk of the second position. The fourth axiom is related to the position size, i.e., the risk proportionally increases with position size. The relevance axiom ensures that if a position always generates negative results (losses), then its risk is positive. The strictness axiom ensures that the measure is sufficiently conservative to exceed the common loss expectation. The law invariance axiom, which is presented for coherent risk measures by Kusuoka (2001), ensures that two positions that have the same probability function have equal risks. This characteristic is important for risk measurement in practice, when real data that are dependent on a law are employed.

**Remark 4**. Given a coherent risk measure $\rho$, Artzner et al. (1999) define the acceptance set as $A_\rho = \{X \in L^p : \rho(X) \leq 0\}$, i.e., the positions that cause a situation with no loss. Let $L^p_+$ be the cone of the non-negative elements of $L^p$ and let $L^p_-$ be its negative counterpart. Each coherent risk measure $\rho$ has an acceptance set $A_\rho$ that satisfies the following properties: contains $L^p_+$, has no intersection with $L^p_-$, is a convex cone. The risk measure associated with this set is $\rho(X) = \inf\{m : X + m \in A_\rho\}$, in other words, the minimum capital that needs to be added to position $X$ to make it acceptable. Artzner et al. (1999) demonstrate that if an acceptance set satisfies the previously defined properties, then the risk measure associated with this set is coherent. If a risk measure is coherent, then the acceptance set linked to this measure satisfies the required properties.

Because the SDR measure is a combination of ES and SD, the first step prior to demonstrating its characteristics is to understand the SD theoretical properties because ES is already well defined in the literature. Because it is a dispersion coefficient, SD is better accommodated within the concept of generalized deviation measures proposed by Rockafellar et al. (2006). We now define generalized deviation measures.

**Definition 3**. *A function $\mathcal{D} : L^p \to \mathbb{R}_+$ is a generalized deviation measure if it fullfils the following axioms:*

> *Translation Insensitivity: $\mathcal{D}(X + C) = \mathcal{D}(X), \forall X \in L^p, C \in \mathbb{R}$*
> *Positive Homogeneity: $\mathcal{D}(\lambda X) = \lambda \mathcal{D}(X), \forall X \in L^p, \lambda \geq 0$.*
> *Subadditivity: $\mathcal{D}(X + Y) \leq \mathcal{D}(X) + \mathcal{D}(Y), \forall X, Y \in L^p$.*
> *Non-Negativity: $\mathcal{D}(X) \geq 0, \forall X \in L^p$, with $\mathcal{D}(X) > 0$ for nonconstant X.*

*In addition, a generalized deviation measure can respect the axioms:*

> *Lower Range Dominance: $\mathcal{D}(X) \leq E_{\mathbb{P}}[X] - \inf X, \forall X \in L^p$*
> *Law Invariance: if $F_X = F_Y$, then $\mathcal{D}(X) = \mathcal{D}(Y), \forall X, Y \in L^p$.*

**Remark 5**. The first axiom indicates that the deviation in relation to the expected value does not change if a constant is added. The second axiom states that the risk of a financial position increases proportionally with its size. The third axiom ensures that the principle of diversification is captured by the measure. The fourth axiom is similar to the relevance concept, which indicates that any nonconstant position exhibits non-negative deviation. Thus, $\mathcal{D}$ captures the degree of uncertainty in $X$ and acts similar to a norm in $L^p$, with the exception that it does not require symmetry. The Lower Range Dominance axiom restricts the deviation measure to a range that is lower than the range between the expected value and the minimum value of the position $X$. The Law Invariance axiom implies that financial positions with the same probability distribution have the same risk as well as guarantees that generalized deviation measures can be estimated from real data.

In addition to the axioms, the continuity properties must be defined because risk measures are basically functions that require these properties to ensure certain results. Thus, we now define continuity properties.

**Definition 4**. Let $\{X_n\}_{n=1}^{\infty} \in L^p$. A risk measure $\rho : L^p \to \mathbb{R}$ is said:

Lipschitz continuous if there is a constant $C \geq 0$ such that $|\rho(X) - \rho(Y)| \leq C\|X - Y\|_p$.

Continuous from above if $X_n \downarrow X$, i.e., $X_n$ converges $\mathbb{P}$ a.s. to $X$ from higher values, implies $\rho(X) = \lim_{n\to\infty} \rho(X_n)$.

Continuous from below if $X_n \uparrow X$, i.e., $X_n$ converges $\mathbb{P}$ a.s. to $X$ from lower values, implies $\rho(X) = \lim_{n\to\infty} \rho(X_n)$.

Fatou continuous if $|X_n| \leq Y$ and $X_n \to X$, i.e., $X_n$ is limited and converges $\mathbb{P}$ a.s. to $X$, with $Y \in L^p$, then $\rho(X) \leq \lim_{n\to\infty} \inf \rho(X_n)$.

Lebesgue continuous if $|X_n| \leq Y$ and $X_n \to X$, i.e., $X_n$ is limited and converges $\mathbb{P}$ a.s. to $X$, with $Y \in L^p$, then $\rho(X) = \lim_{n\to\infty} \rho(X_n)$.

Given the continuity properties, a coherent risk measure can be represented as the worst possible expected result of $X$ from the scenarios generated by the probability measures $\mathbb{Q} \in \mathcal{P}$. Artzner et al. (1999) present this result for discrete $L^\infty$ spaces. Delbaen (2002) generalizes for continuous $L^\infty$ spaces, and Inoue (2003) considers the spaces $L^p, 1 \leq p < \infty$. It is also possible to represent generalized deviation measures in a similar approach, with the due adjustments, as demonstrated by Rockafellar et al. (2006). These representations are formally guaranteed by the following results.

**Theorem 1**. Delbaen (2002), Inoue (2003). $\rho : L^p \to \mathbb{R}$ is a Fatou continuous coherent risk measure if and only if it can be represented in accordance with $\rho(X) = \sup_{\mathbb{Q} \in \mathcal{P}_\rho} E_{\mathbb{Q}}[-X]$, where $\mathcal{P}_\rho$ is a closed and convex subset of $\mathcal{P}$ and $\mathcal{P}_\rho = \{\mathbb{Q} \in \mathcal{P} : \frac{d\mathbb{Q}}{d\mathbb{P}} \in L^q, \frac{1}{p} + \frac{1}{q} = 1, \frac{d\mathbb{Q}}{d\mathbb{P}} \geq 0, E_{\mathbb{P}}\left[\frac{d\mathbb{Q}}{d\mathbb{P}}\right] = 1\}$.

**Theorem 2**. Rockafellar et al. (2006). A function $\mathcal{D} : L^p \to \mathbb{R}_+$ is a Fatou continuous generalized deviation measure if and only if it can be represented as $\mathcal{D}(X) = E_{\mathbb{P}}[X] - \inf_{\mathbb{Q} \in \mathcal{P}_\mathcal{D}} E_{\mathbb{Q}}[X]$, where $\mathcal{P}_\mathcal{D}$ is a closed and convex subset of $\mathcal{P}$, where for any no-constant $X$ there is $\mathbb{Q} \in \mathcal{P}_\mathcal{D}$ with $E_\mathbb{Q}[X] < E_\mathbb{P}[X]$, such that $\mathcal{P}_\mathcal{D} = \left\{ \mathbb{Q} \in \mathcal{P} : \frac{d\mathbb{Q}}{d\mathbb{P}} \in L^q, \frac{1}{p} + \frac{1}{q} = 1, E_\mathbb{P}\left[\frac{d\mathbb{Q}}{d\mathbb{P}}\right] = 1, \mathcal{D}(X) \geq E_\mathbb{P}[X] - E_\mathbb{Q}[X], \forall X \in L^p \right\}$. The finiteness of $\mathcal{D}$ is equivalent to the boundedness of $\mathcal{P}_\mathcal{D}$ . $\mathcal{D}$ satisfies the Lower Range Dominance axiom if and only if $\frac{d\mathbb{Q}}{d\mathbb{P}} \geq 0, \forall \mathbb{Q} \in \mathcal{P}_\mathcal{D}$. The set $\mathcal{P}_\mathcal{D}$ uniquely defined for a generalized deviation measure $\mathcal{D}$ is referred to as the risk envelope.

### 3. Main results

The main purpose of this section is to demonstrate the theoretical properties of SDR as a risk measure. Based on definitions and results from the previous section, we first prove the axioms and representation of the SD as a generalized deviation measure.

**Theorem 3**. *The function $SD^\alpha : L^p \to \mathbb{R}_+$ defined in (3) is a generalized deviation measure that satisfies the Lower Range Dominance and Law Invariance axioms, with risk envelope $\mathcal{P}_{SD^\alpha}$, where*

$$\mathcal{P}_{SD^\alpha} = \left\{ \mathbb{Q} \in \mathcal{P} : \frac{d\mathbb{Q}}{d\mathbb{P}} \in L^q, \frac{1}{p} + \frac{1}{q} = 1, \frac{d\mathbb{Q}}{d\mathbb{P}} \geq 0, E_\mathbb{P}\left[\frac{d\mathbb{Q}}{d\mathbb{P}}\right] = 1, \frac{SD^\alpha(X)}{\sigma(X)} \geq \sigma\left(\frac{d\mathbb{Q}}{d\mathbb{P}} - 1\right), \forall X \in L^p \right\}.$$

*Proof.* First, it is necessary to prove that SD satisfies the axioms. Thus, we prove them one by one. Based on the axioms and the previous results, we obtain the dual representation.

i) *Translation Insensitivity*. Because $e_\alpha(X + C) = e_\alpha(\text{X}) + C$, we have

$$\begin{aligned} SD^\alpha(X + C) &= \left\| (X + C - e_\alpha(X + C))^- \right\|_p \\ &= \left\| (X + C - e_\alpha(X) - C)^- \right\|_p \\ &= \left\| (X - e_\alpha(X))^- \right\|_p \\ &= SD^\alpha(X). \end{aligned}$$

ii) *Positive Homogeneity*. Because $e_\alpha(\lambda X) = \lambda e_\alpha(X)$ for $\lambda \geq 0$, we have

$$\begin{aligned} SD^\alpha(\lambda X) &= \left\| (\lambda X - e_\alpha(\lambda X))^- \right\|_p \\ &= \left\| \lambda (X - e_\alpha(X))^- \right\|_p \\ &= \lambda \left\| (X - e_\alpha(X))^- \right\|_p \\ &= \lambda SD^\alpha(X). \end{aligned}$$

iii) *Subadditivity*. Because $\|X\|_p$, $(X)^-$ and $ES^\alpha(X) = -e_\alpha(X)$ are subbaditive, we have

$$\begin{aligned} SD^\alpha(X + Y) &= \left\| (X + Y - e_\alpha(X + Y))^- \right\|_p \\ &\leq \left\| (X - e_\alpha(X) + Y - e_\alpha(Y))^- \right\|_p \\ &\leq \left\| (X - e_\alpha(X))^- + (Y - e_\alpha(Y))^- \right\|_p \\ &\leq \left\| (X - e_\alpha(X))^- \right\|_p + \left\| (Y - e_\alpha(Y))^- \right\|_p \\ &= SD^\alpha(X) + SD^\alpha(Y). \end{aligned}$$

iv) *Non-Negativity*. By definition, SD is a p-norm that can only assume non-negative values. For strict positivity in the case of a nonconstant $X$, noted that given the set of values $x = \{X_i \in X : X_i \leq e_\alpha(X)\}$, $x$ will not be constant if $\exists X_i \in x : X_i \neq e_\alpha(X)$. Thus, $\exists X_i \in x : (X_i - e_\alpha(X))^- > 0$, and, by its definition, $SD^\alpha(X) > 0$.

v) *Lower Range Dominance*. Consider the sequence of inequalities $E_{\mathbb{P}}[X] - \inf X \geq e_\alpha(X) - \inf X \geq \left(X - e_\alpha(X)\right)^-$. Because $\left(E_{\mathbb{P}}[(C)^p]\right)^{\frac{1}{p}} = C$ for constant $C \geq 0$ and that performing these operations on both sides does not change the inequalities because both terms are non-negative, we have $e_\alpha(X) - \inf X \geq \left(E_{\mathbb{P}}\left[\left|\left(X - e_\alpha(X)\right)^-\right|^p\right]\right)^{\frac{1}{p}} = SD^\alpha(X)$. Thus, $SD^\alpha(X) \leq E_{\mathbb{P}}[X] - \inf X$.

vi) *Law Invariance*. Assuming that $F_X = F_Y$, we have

$$SD^\alpha(X) = \left(E_{\mathbb{P}}\left[\left|\left(X - e_\alpha(X)\right)^-\right|^p\right]\right)^{\frac{1}{p}}$$
$$= \left(E_{\mathbb{P}}\left[\left|\left(Y - e_\alpha(Y)\right)^-\right|^p\right]\right)^{\frac{1}{p}}$$
$$= SD^\alpha(Y).$$

Thus, $SD^\alpha(X) = SD^\alpha(Y)$ for $F_X = F_Y$.

vii) *Dual representation*. SD is convex because it complies with the Subadditivity and Positive Homogeneity axioms. It also satisfies the Law Invariance axiom, which implies that it is Fatou continuous, as demonstrated by Jouini et al. (2006) for an atomless space. According to Theorem 2, SD can be characterized by dual representation. The conjugate space of $L^p$ is $L^q$ with $\frac{1}{p} + \frac{1}{q} = 1$. The risk envelope formed by the probability measures $\mathbb{Q} \in \mathcal{P}$ should be defined such that

$$SD^\alpha(X) \geq E_{\mathbb{P}}[X] - E_{\mathbb{Q}}[X]$$
$$= E_{\mathbb{P}}\left[X\left(1 - \frac{d\mathbb{Q}}{d\mathbb{P}}\right)\right]$$
$$= E_{\mathbb{P}}[X]E_{\mathbb{P}}\left[1 - \frac{d\mathbb{Q}}{d\mathbb{P}}\right] + \sigma(X)\sigma\left(1 - \frac{d\mathbb{Q}}{d\mathbb{P}}\right)cor\left(X, 1 - \frac{d\mathbb{Q}}{d\mathbb{P}}\right)$$
$$= \sigma(X)\sigma\left(\frac{d\mathbb{Q}}{d\mathbb{P}} - 1\right)$$

Such steps are justified because $E_{\mathbb{P}}[XY] = E_{\mathbb{P}}[X]E_{\mathbb{P}}[Y] + \sigma(X)\sigma(Y)cor(X,Y)$ and $cor\left(X, M(X)\right) = 1$, where $M(X)$ is a direct function of $X$. Moreover, according to Theorem 2, we have $E_{\mathbb{P}}\left[\frac{d\mathbb{Q}}{d\mathbb{P}}\right] = 1$ and $\frac{d\mathbb{Q}}{d\mathbb{P}} \geq 0$; this last inequality due to the Lower Range Dominance axiom. Thus, $\frac{SD^\alpha(X)}{\sigma(X)} \geq \sigma\left(\frac{d\mathbb{Q}}{d\mathbb{P}} - 1\right)$. Therefore, we have the dual representation $SD^\alpha(X) = E_{\mathbb{P}}[X] - \inf_{\mathbb{Q} \in \mathcal{P}_{SD^\alpha}} E_{\mathbb{Q}}[X]$, where $\mathcal{P}_{SD^\alpha} = \left\{\mathbb{Q} \in \mathcal{P} : \frac{d\mathbb{Q}}{d\mathbb{P}} \geq 0, \frac{d\mathbb{Q}}{d\mathbb{P}} \in L^q, E_{\mathbb{P}}\left[\frac{d\mathbb{Q}}{d\mathbb{P}}\right] = 1, \frac{SD^\alpha(X)}{\sigma(X)} \geq \sigma\left(\frac{d\mathbb{Q}}{d\mathbb{P}} - 1\right), \forall X \in L^p\right\}$. $\square$

**Remark 6**. Grechuk et al. (2009) show that a generalized deviation measure with the Law Invariance axiom can be represented as $\mathcal{D}(X) = \sup_{\phi(\alpha) \in \Lambda} \int_0^1 (ES^\alpha(X) - E_{\mathbb{P}}[X]) \, d\left(\psi(\alpha)\right)$, where $\Lambda$ is a set of nonincreasing functions $\phi(\alpha) \in L^q, \frac{1}{p} + \frac{1}{q} = 1, \int_0^1 \phi(\alpha) \, d\alpha = 0, \phi(\alpha) = q_\alpha\left(\frac{d\mathbb{Q}}{d\mathbb{P}}\right) = \frac{1}{\alpha}\int_0^\alpha \psi(d\alpha)$. This representation is inspired in coherent measures with the Law Invariance axiom and spectral measures, which were proposed by Kusuoka (2001) and Acerbi (2002), respectively. In the case of SD, we have $\Lambda = \left\{\phi(\alpha) : \phi(\alpha) \in L^2, \int_0^1 \phi(\alpha) \, d\alpha = 0, \phi(\alpha) = q_\alpha\left(\frac{d\mathbb{Q}}{d\mathbb{P}}\right), 1 - \mathbb{Q} \in \mathcal{P}_{SD^\alpha}\right\}$.

Based on the properties proven for SD, we turn our focus now to SDR. Because SDR is a combination of ES and SD, we use known properties of the two measures to prove the SDR axioms. We also discuss issues such as representations and implications regarding other theoretical results in the literature.

**Theorem 4**. *The function $SDR^\alpha : L^p \to \mathbb{R}$ defined in (4) is a coherent risk measure that satisfies the Relevance, Strictness and Law Invariance axioms. In addition,* $\sup -X \geq SDR^\alpha \geq ES^\alpha \geq VaR^\alpha$ *and $SDR^\alpha$ is non-increasing in $\alpha$ and $\beta$. Its dual representation is* $SDR^\alpha = \sup\limits_{\mathbb{Q} \in \mathcal{P}_{SDR^\alpha}} \{E_\mathbb{Q}[-X]\},$ *where*

$\mathcal{P}_{SDR^\alpha} = \left\{ \mathbb{Q} \in \mathcal{P} : \frac{d\mathbb{Q}}{d\mathbb{P}} = \frac{d\mathbb{Q}_1}{d\mathbb{P}} + \frac{d\mathbb{Q}_2}{d\mathbb{P}} - 1, \mathbb{Q}_1 \in \mathcal{P}_{ES^\alpha}, \mathbb{Q}_2 \in \mathcal{P}_{(1-\alpha)^\beta SD^\alpha} \right\};$

$\mathcal{P}_{(1-\alpha)^\beta SD^\alpha} = \left\{ \mathbb{Q} \in \mathcal{P} : \frac{d\mathbb{Q}}{d\mathbb{P}} = \left[ 1 - (1-\alpha)^\beta \right] + (1-\alpha)^\beta \frac{d\mathbb{Q}'}{d\mathbb{P}}, \frac{d\mathbb{Q}'}{d\mathbb{P}} \in \mathcal{P}_{SD^\alpha} \right\};$

$\mathcal{P}_{ES^\alpha} = \left\{ \mathbb{Q} \in \mathcal{P} : \frac{d\mathbb{Q}}{d\mathbb{P}} \geq 0, E_\mathbb{P} \left[ \frac{d\mathbb{Q}}{d\mathbb{P}} \right] = 1, \frac{d\mathbb{Q}}{d\mathbb{P}} \leq \frac{1}{\alpha} \right\}.$

*Proof.* First, it is necessary to prove that SDR satisfies the axioms. Based on the axioms and previous results, we demonstrate the relationship with the other measures, its non-increasingness in the parameters and the dual representation.

iv) *Translation Invariance*. Because ES satisfies this axiom and SD satisfies the Translation Insensitivity axiom, we have

$SDR^\alpha(X + C) = ES^\alpha(X + C) + (1-\alpha)^\beta SD^\alpha(X + C)$
$\qquad\qquad = ES^\alpha(X) + (1-\alpha)^\beta SD^\alpha(X) - C = SDR^\alpha(X) - C.$

ii) *Subadditivity*. Because both ES and SD are subadditive, $SDR^\alpha(X + Y) = ES^\alpha(X + Y) + (1-\alpha)^\beta SD^\alpha(X + Y)$
$\qquad\qquad \leq ES^\alpha(X) + ES^\alpha(Y) + (1-\alpha)^\beta [SD^\alpha(X) + SD^\alpha(Y)]$
$\qquad\qquad = ES^\alpha(X) + (1-\alpha)^\beta SD^\alpha(X) + ES^\alpha(Y) + (1-\alpha)^\beta SD^\alpha(Y)$
$\qquad\qquad = SDR^\alpha(X) + SDR^\alpha(Y)$

iii) *Monotonicity*. Assuming that $X \leq Y, Z \geq 0$, such that $X + Z = Y$. Due to the Lower Range Dominance axiom of SD, $Z \geq 0$ and $0 \leq (1-\alpha)^\beta \leq 1$, we have $(1-\alpha)^\beta SD^\alpha(Z) \leq -ES^\alpha(Z)$. Thus, $SDR^\alpha(Z) \leq 0$. According to the SDR Subadditivity, we have $SDR^\alpha(Y) = SDR^\alpha(X + Z) \leq SDR^\alpha(X) + SDR^\alpha(Z) \leq SDR^\alpha(X)$.

iv) *Positive Homogeneity*. Because both ES and SD have this axiom, for $\lambda \geq 0$, we have

$SDR^\alpha(\lambda X) = ES^\alpha(\lambda X) + (1-\alpha)^\beta SD^\alpha(\lambda X)$
$\qquad\qquad = \lambda[ES^\alpha(X) + (1-\alpha)^\beta SD^\alpha(X)] = \lambda SDR^\alpha(X).$

v) *Relevance*. Because ES respects this axiom, as it is an expectation, and SD satisfies the Non-Negativity axiom, for $X \leq 0$ and $X \neq 0$, we have $0 < ES^\alpha(X) \leq ES^\alpha(X) + (1-\alpha)^\beta SD^\alpha(X) = SDR^\alpha(X)$. Thus, $SDR^\alpha(X) > 0$.

vi) *Strictness*. By definition, $ES^1 = -E_\mathbb{P}[X]$. Because ES is decreasing in $\alpha$ and the SD satisfies the Non-Negativity axiom, we have $-E_\mathbb{P}[X] \leq ES^\alpha \leq ES^\alpha(X) + (1-\alpha)^\beta SD^\alpha(X) = SDR^\alpha(X)$. Thus, $SDR^\alpha(X) \geq -E_\mathbb{P}[X]$.

vii) *Law Invariance*. Because ES and SD satisfy this axiom, assuming that $F_X = F_Y$, we have $SDR^\alpha(X) = ES^\alpha(X) + (1-\alpha)^\beta SD^\alpha(X) = ES^\alpha(Y) + (1-\alpha)^\beta SD^\alpha(Y) = SDR^\alpha(Y).$

viii) *Relationship with other measures*. As ES is an expectation that considers the information beyond VaR, $ES^\alpha(X) \geq VaR^\alpha(X)$. Because SD satisfies the Non-Negativity axiom, we have $SDR^\alpha(X) = ES^\alpha(X) + (1-\alpha)^\beta SD^\alpha(X) \geq ES^\alpha(X)$. According to the Lower Range Dominance axiom, $SDR^\alpha(X) = ES^\alpha(X) + (1-\alpha)^\beta SD^\alpha(X) \leq -\inf X = \sup -X$. Thus, $\sup -X \geq SDR^\alpha(X) \geq ES^\alpha(X) \geq VaR^\alpha(X)$.

ix) *Nonincreasing in parameters*. From $(1-\alpha) \leq 1$ and $\beta \geq 0$, we have that $(1-\alpha)^\beta$ is non-increasing in $\beta$. To prove that is same case for $\alpha$, assume the opposite, i.e., $SDR^{\alpha_1}(X) > SDR^{\alpha_2}(X)$ for $\alpha_1 \geq \alpha_2$. Thus, we have $SDR^\alpha(X) > SDR^0(X)$ because $\alpha \geq 0$. However, $SDR^0(X) = ES^0(X) + (1-\alpha)^\beta SD^0(X) = -\inf X$, implying that $SDR^\alpha(X) > -\inf X$, but this is a contradiction because $SDR^\alpha(X) \leq -\inf X = \sup -X$.

x) *Dual representation*. Because SDR is coherent and satisfies the Law Invariance axiom, it is Fatou continuous, as demonstrated in Jouini et al. (2006). According to Theorem 1, we can characterize the SDR according to the dual representation. We have to define the subset formed by the probability measures $\mathbb{Q} \in \mathcal{P}$. According to Rockafellar et al. (2006), generalized deviation measures such as $\lambda \mathcal{D}(X)$, with $\lambda \geq 0$, have the risk envelope $\mathcal{P}_{\lambda \mathcal{D}} = \left\{ \mathbb{Q} \in \mathcal{P} : \frac{d\mathbb{Q}}{d\mathbb{P}} = (1-\lambda) + \lambda \frac{d\mathbb{Q}'}{d\mathbb{P}}, \frac{d\mathbb{Q}'}{d\mathbb{P}} \in \mathcal{P}_{\mathcal{D}} \right\}$. Thus, $\mathcal{P}_{(1-\alpha)^\beta SD^\alpha} = \left\{ \mathbb{Q} \in \mathcal{P} : \frac{d\mathbb{Q}}{d\mathbb{P}} = \left[ 1 - (1-\alpha)^\beta \right] + (1-\alpha)^\beta \frac{d\mathbb{Q}'}{d\mathbb{P}}, \frac{d\mathbb{Q}'}{d\mathbb{P}} \in \mathcal{P}_{SD^\alpha} \right\}$. Delbaen (2002) shows that $ES^\alpha(X) = \sup\limits_{\mathbb{Q} \in \mathcal{P}_{ES^\alpha}} E_{\mathbb{Q}}[-X]$, $\mathcal{P}_{ES^\alpha} = \left\{ \mathbb{Q} \in \mathcal{P} : \frac{d\mathbb{Q}}{d\mathbb{P}} \geq 0, E_{\mathbb{P}} \left[ \frac{d\mathbb{Q}}{d\mathbb{P}} \right] = 1, \frac{d\mathbb{Q}}{d\mathbb{P}} \leq \frac{1}{\alpha} \right\}$. Therefore, we have

$$SDR^\alpha(X) = ES^\alpha(X) + (1-\alpha)^\beta SD^\alpha(X)$$
$$= \sup\limits_{\mathbb{Q}_1 \in \mathcal{P}_{ES^\alpha}} E_{\mathbb{Q}_1}[-X] + E_{\mathbb{P}}[X] - \inf\limits_{\mathbb{Q}_2 \in \mathcal{P}_{(1-\alpha)^\beta SD^\alpha}} E_{\mathbb{Q}_2}[X]$$
$$= \sup\limits_{\mathbb{Q}_1 \in \mathcal{P}_{ES^\alpha}, \mathbb{Q}_2 \in \mathcal{P}_{(1-\alpha)^\beta SD^\alpha}} \left\{ E_{\mathbb{Q}_1}[-X] - E_{\mathbb{P}}[-X] + E_{\mathbb{Q}_2}[-X] \right\}$$
$$= \sup\limits_{\mathbb{Q}_1 \in \mathcal{P}_{ES^\alpha}, \mathbb{Q}_2 \in \mathcal{P}_{(1-\alpha)^\beta SD^\alpha}} \left\{ E_{\mathbb{P}} \left[ -X \left( \frac{d\mathbb{Q}_1}{d\mathbb{P}} + \frac{d\mathbb{Q}_2}{d\mathbb{P}} - 1 \right) \right] \right\}$$
$$= \sup\limits_{\mathbb{Q} \in \mathcal{P}_{SDR^\alpha}} \left\{ E_{\mathbb{Q}}[-X] \right\}.$$

where $\mathcal{P}_{SDR^\alpha} = \left\{ \mathbb{Q} \in \mathcal{P} : \frac{d\mathbb{Q}}{d\mathbb{P}} = \frac{d\mathbb{Q}_1}{d\mathbb{P}} + \frac{d\mathbb{Q}_2}{d\mathbb{P}} - 1, \mathbb{Q}_1 \in \mathcal{P}_{ES^\alpha}, \mathbb{Q}_2 \in \mathcal{P}_{(1-\alpha)^\beta SD^\alpha} \right\}$. To show that $\mathcal{P}_{SDR^\alpha}$ is composed by valid measures, one must only verify that for $\mathbb{Q} \in \mathcal{P}_{SDR^\alpha}$, $\mathbb{Q}_1 \in \mathcal{P}_{ES^\alpha}$, $\mathbb{Q}_2 \in \mathcal{P}_{(1-\alpha)^\beta SD^\alpha}$, $E_{\mathbb{P}} \left[ \frac{d\mathbb{Q}}{d\mathbb{P}} \right] = E_{\mathbb{P}} \left[ \frac{d\mathbb{Q}_1}{d\mathbb{P}} \right] + E_{\mathbb{P}} \left[ \frac{d\mathbb{Q}_2}{d\mathbb{P}} \right] - E_{\mathbb{P}}[1] = 1$. In addition $\frac{d\mathbb{Q}}{d\mathbb{P}} \geq 0$ because assuming the opposite would yield $\frac{d\mathbb{Q}}{d\mathbb{P}} < 0$, and therefore, $E_{\mathbb{P}} \left[ \frac{d\mathbb{Q}}{d\mathbb{P}} \right] < 0$. Thus, $2 = E_{\mathbb{P}} \left[ \frac{d\mathbb{Q}_1}{d\mathbb{P}} \right] + E_{\mathbb{P}} \left[ \frac{d\mathbb{Q}_2}{d\mathbb{P}} \right] < E_{\mathbb{P}}[1] = 1$, a contradiction. □

**Remark 7**. The Law Invariance axiom is fundamental because it implies that the risk measure can be estimated by real data; in other words, it can be employed for practical risk measurement. Kusuoka (2001) shows that a coherent risk measure that satisfies the Law Invariance axiom and is Fatou continuous can be mathematically represented as $\rho(X) = \sup\limits_{m \in \mathcal{P}_{(0,1]}} \int_0^1 ES^\alpha(X) m(d\alpha)$, where $\mathcal{P}_{(0,1]}$ are probability measures defined in $(0,1]$. Jouini et al. (2006) show that law invariant convex risk measures that are defined in standard spaces will automatically be Fatou continuous. Svindland (2010) generalizes the results of Jouini et al.

(2006) by relaxing the assumption that the probability space is standard and only requiring it to be atomless. Because the SDR is coherent and satisfies the Law Invariance axiom, it can be represented as that supreme of ES combinations. As we are considering an atomless space, we can define a continuous variable $U \sim \mathbb{U}(0,1)$, uniformly distributed between 0 and 1, such that $F_X^{-1}(U) = X$. For $\mathbb{Q} \in \mathcal{P}_{SDR^\alpha}$, we can represent $\frac{d\mathbb{Q}}{d\mathbb{P}} = H(U)$, where $H$ is a monotonically decreasing function. To obtain the supreme in a dual representation, $\frac{d\mathbb{Q}}{d\mathbb{P}}$ must be anti-monotonic in relation to $X$. Letting $H(U) = \int_{(u,1]} \frac{1}{\alpha} dm(\alpha)$ with $m \in \mathcal{P}_{(0,1]}$ and knowing that $ES^\alpha(X) = -\frac{1}{\alpha} \int_0^\alpha F_X^{-1}(u) du$ for continuous distributions, based on the dual representation, we have

$$
\begin{aligned}
SDR^\alpha(X) &= \sup_{\mathbb{Q} \in \mathcal{P}_{SDR^\alpha}} \{E_\mathbb{Q}[-X]\} = \sup_{\mathbb{Q} \in \mathcal{P}_{SDR^\alpha}} \left\{ E_\mathbb{P}\left[ -X \frac{d\mathbb{Q}}{d\mathbb{P}} \right] \right\} \\
&= \sup_{m \in \mathcal{P}_{(0,1]}} \left\{ \int_0^1 -F_X^{-1}(u) \left[ \int_{(u,1]} \frac{1}{\alpha} dm(\alpha) \right] du \right\} \\
&= \sup_{m \in \mathcal{P}_{(0,1]}} \left\{ \int_{(0,1]} \left[ \frac{1}{\alpha} \int_0^\alpha -F_X^{-1}(u) du \right] dm(\alpha) \right\} \\
&= \sup_{m \in \mathcal{P}_{(0,1]}} \left\{ \int_{(0,1]} ES^\alpha(X) dm(\alpha) \right\}.
\end{aligned}
$$

Thus, we only must to define the measures $m \in \mathcal{P}_{(0,1]}$ that are candidates. By the same logic, we have $\frac{d\mathbb{Q}_1}{d\mathbb{P}} = H_1(u) = \int_{(u,1]} \frac{1}{\alpha} dm_1(\alpha)$, $\frac{d\mathbb{Q}_2}{d\mathbb{P}} = H_2(u) = \int_{(u,1]} \frac{1}{\alpha} dm_2(\alpha)$ and $\frac{d\mathbb{P}}{d\mathbb{P}} = H_3(u) = \int_{(u,1]} \frac{1}{\alpha} dm_3(\alpha) = 1$. Thus, by converting the restrictions, the following sets are obtained:

$M_1 = \left\{ m_1 \in \mathcal{P}_{(0,1]} : \int_{(u,1]} \frac{1}{\alpha} dm_1(\alpha) \geq 0, \int_{(0,1]} dm_1(u) = 1, \int_{(u,1]} \frac{1}{\alpha} dm_1(\alpha) \leq \frac{1}{\alpha} \right\}$;

$M_2 = \left\{ m_2 \in \mathcal{P}_{(0,1]} : \int_{(u,1]} \frac{1}{\alpha} dm_2(\alpha) = \left[ 1 - (1-\alpha)^\beta \right] + (1-\alpha)^\beta \int_{(u,1]} \frac{1}{\alpha} dm_{2'}(\alpha), m_{2'} \in M_{2'} \right\}$;

$M_{2'} = \left\{ m_{2'} \in \mathcal{P}_{(0,1]} : \int_{(u,1]} \frac{1}{\alpha} dm_{2'}(\alpha) \geq 0, \int_{(u,1]} \frac{1}{\alpha} dm_{2'}(\alpha) \in L^q, \int_{(0,1]} dm_1(u) = 1, \frac{SD^\alpha(X)}{\sigma(X)} \geq \sigma\left( \int_{(u,1]} \frac{1}{\alpha} dm_{2'}(\alpha) - 1 \right), \forall X \in L^p \right\}$;

$M_3 = \left\{ m_3 \in \mathcal{P}_{(0,1]} : \int_{(u,1]} \frac{1}{\alpha} dm_3(\alpha) = 1 \right\}$.

Because $\frac{d\mathbb{Q}}{d\mathbb{P}} = \frac{d\mathbb{Q}_1}{d\mathbb{P}} + \frac{d\mathbb{Q}_2}{d\mathbb{P}} - 1$, $\mathbb{Q}_1 \in \mathcal{P}_{ES^\alpha}$, $\mathbb{Q}_2 \in \mathcal{P}_{(1-\alpha)^\beta SD^\alpha}$, we have $\frac{d\mathbb{Q}}{d\mathbb{P}} = H(u) = \int_{(u,1]} \frac{1}{\alpha} dm(\alpha)$, with $dm(\alpha) = dm_1(\alpha) + dm_2(\alpha) - dm_3(\alpha)$. Thus, we obtain the expression $SDR^\alpha(X) = \sup_{m \in M} \left\{ \int_{(0,1]} ES^\alpha(X) dm(\alpha) \right\}$, where $M = \{ m \in \mathcal{P}_{(0,1]} : dm(\alpha) = dm_1(\alpha) + dm_2(\alpha) - dm_3(\alpha), m_1 \in M_1, m_2 \in M_2, m_3 \in M_3 \}$.

**Remark 8**. Because SDR is a coherent risk measure, an acceptance set that contains $L_+^p$, with no intersection with $L_-^p$ and that is a convex cone, can be established according to Artzner et al. (1999). This set assumes the form $A_{SDR^\alpha} = \{ X \in L^p : SDR^\alpha(X) \leq 0 \}$, with $SDR^\alpha(X) = \inf\{ m : X + m \in A_{SDR^\alpha} \}$. This concept is closely linked to the issue of capital regulation because it enables checking the amount of funds that need to be maintained to prevent the loss measured by SDR, i.e., to make the position acceptable because $SDR^\alpha(X + SDR^\alpha(X)) = 0$. If the value is negative, it represents the amount of capital that can be withdrawn without the position becoming unacceptable. The acceptance set logic is derived from the Translation Invariance axiom. El Karoui and Ravanelli (2009) relax the Translation Invariance axiom to a subadditive form to address uncertainty in interest rates. This axiom states that $\rho(X - C) \leq \rho(X) + C, \forall X \in L^p, C \in \mathbb{R}$. Obviously, the Translation Invariance axiom is a special case, in the way that the SDR satisfies this axiom in the subadditive form.

**Remark 9**. Because it is coherent, SDR is included in the class of convex risk measures proposed by Föllmer and Schied (2002) and Frittelli and Gianin (2002). This class relaxes the Positive Homogeneity and Subadditivity axioms and substitutes them with the weaker Convexity axiom. This axiom specifies that the risk of a diversified position is less than or equal to the weighted mean of individual risks, which in mathematical form gives us $\rho(\lambda X + (1 - \lambda)Y) \leq \lambda \rho(X) + (1 - \lambda)\rho(Y), \forall X, Y \in L^p, 0 \leq \lambda \leq 1$. Regarding dual representation, convex risk measures are defined according to $\rho(X) = \sup_{\mathbb{Q} \in \mathcal{P}_\rho} (E_P[-X] - \alpha(\mathbb{Q}))$, where $\alpha : \mathcal{P} \to (-\infty, \infty]$ is a convex penalty function that is continuous from below, according to $\alpha(\mathbb{Q}) = \sup_{X \in A_\rho} (E_\mathbb{Q}[-X])$, with $\alpha(\mathbb{Q}) \geq -\rho(0)$. For the SDR case, we have $SDR^\alpha(X) = \sup_{\mathbb{Q} \in \mathcal{P}_{SDR^\alpha}} \{E_\mathbb{Q}[-X] - \alpha(\mathbb{Q})\}, \quad \alpha(\mathbb{Q}) = \sup_{X \in A_{SDR^\alpha}} (E_\mathbb{Q}[-X]) = 0$. Convexity is crucial for optimization problems, such as all problems related to resource allocation. Thus, SDR is a valid measure to be considered within resource allocation. Moreover, it would be an interesting enhancement to the development of SDR as a solution to an optimization problem.

**Remark 10**. Because SDR is a function, continuity properties become interesting. As noted in the proof of Theorem 4, because it is convex and satisfies the Law Invariance axiom, the result of Jouini et al. (2006) ensures that SDR is Fatou continuous. Due to the Translation Invariance and Monotonicity axioms, $Y \leq X + \|Y - X\|_p$ implies $SDR^\alpha(Y) \geq SDR^\alpha(X) - \|Y - X\|_p$. Thus, $SDR^\alpha(X) - SDR^\alpha(Y) \leq \|Y - X\|_p$. By inverting the roles of $X$ and $Y$, we have $SDR^\alpha(Y) - SDR^\alpha(X) \leq \|X - Y\|_p$. By combining the two inequalities, $|SDR^\alpha(X) - SDR^\alpha(Y)| \leq \|X - Y\|_p$. Hence, the SDR is Lipschitz continuous. According to Krätschmer (2005), SDR is continuous from above because it fits in the class of convex measures and has dual representation. As we only consider finite values, SDR is continuous from below and Lebesgue continuous, according to the results proven in Kaina and Rüschendorf (2009) for finite convex measures.

**Remark 11**. A key issue is the use of risk measures for decision-making. Therefore, it is very important that risk measures respect stochastic dominance orders. Based on Leitner (2005) and Bäuerle and Müller (2006), SDR, by satisfying the Law Invariance, Monotonicity and Convexity axioms, complies with second-order stochastic dominance; in other words, $X \preceq_{2std} Y$ implies $\rho(X) \geq \rho(Y)$. Thus, investors who are risk-averse, i.e., with concave utility functions, have their preferences reflected by the SDR. Thus, law-invariant convex risk measures have the property that $\rho(E_\mathbb{P}[X|\mathcal{G}]) \leq \rho(X)$, the risk of a position is greater than the risk of its conditional expected value at any $\mathcal{G} \subseteq \mathcal{F}$. This relationship is related to the modification of the Monotonicity axiom, which is the Dilated Monotonicity axiom introduced by Leitner (2004) for Fatou continuous coherent risk measures. Consider $\mathcal{G} \subseteq \tilde{\mathcal{G}}$, where $\tilde{\mathcal{G}}$ is the family of all possible event subspaces $\mathcal{F}$. $Y$ is a dilation of $X$, $X \preceq_\mathcal{G}^b Y$, if there is a $\tilde{\mathcal{F}} \in \mathcal{G}$ such that $E_\mathbb{P}[X|\tilde{\mathcal{F}}] \leq Y$. If the measure satisfies the Dilated Monotonicity axiom, it is guaranteed that if $X \preceq_\mathcal{G}^b Y$, then $\rho(X) \geq \rho(Y), \forall X, Y \in L^p$. SDR satisfies this axiom because, as demonstrated by Cherny and Grigoriev (2007), every convex risk measure satisfying the Law Invariance axiom, which is defined in atomless probability space, also satisfies the Dilated Monotonicity axiom. Based on these properties, SDR becomes a very interesting measure for decision-making applications because it reflects preference relations.

**Remark 12**. SDR also fits into other flexible risk measure classes, such as the natural risk measures introduced by Kou et al. (2013). The axioms satisfied by this class of measures, in addition to the Positive Homogeinity and Monotonicity axioms, include the Translation Scalling axiom (a scaled version of the Translation Invariance axiom), the Comonotonic Subadditivity axiom (subadditivity only required for comonotonic variables, that is, with perfect positive association) and the Empirical Law Invariance axiom (a version that includes any permutation in the data). Because these axioms are weaker than the axioms that we prove for the SDR, the proposed measure is part of this class. Moreover, if we consider the adaptation $SDR^{*,\alpha}(X) = SDR^\alpha(\min(X, 0))$, the resulting measure is included in the definitions of Cont et al. (2010) and Staum (2013), who introduced measures that only consider losses, with the Monotonicity and Convexity axioms and an adaptation of the Translation Invariance axiom.

In addition to an intuitive financial and economic meaning, SDR possesses solid theoretical properties. Based on this structure, we argue that SDR is an important risk measure for use in financial problems, such as practical risk measurement, capital requirement, resource allocation and decision-making, as well as other areas of knowledge.

### 4. Illustrations

In this section, we present an illustration using simulations and real data to explore such concepts in a more practical way. We expose some plots, based on simple simulations, to visualize the definition of the measure. In addition, results based on Monte Carlo simulation and real data are employed to illustrate the relationship with the predominant risk measures, VaR and ES, for different scenarios and periods. The focus is not the analysis of issues such as modeling, backtesting or even details on different financial applications, but the behavior of SDR when applied to financial data.

We use the empirical method, known as historical simulation (HS), which is a nonparametric method that creates no assumptions about the data and is the most extensively used method in academic studies and in the financial industry. Pérignon and Smith (2010) indicate that 76% of institutions that disclose their risk estimation procedures use HS. Although HS has been criticized (Pritsker, 2006), the focus here is not to discuss estimation details or even compare models. Thus, we choose the most common, simple and flexible model. More specifically, let $F_X^E$ be the empirical distribution of $X$; then estimators of the measures considered are conforming (5):

$$\widehat{VaR}^\alpha = -(F_X^E)^{-1}(\alpha),$$
$$\widehat{ES}^\alpha = -(N\alpha)^{-1} \sum_{i=1}^N \left( \{X\}_1^N * \mathbf{1}_{\{X\}_1^N < -\widehat{VaR}^\alpha} \right),$$
$$\widehat{SD}^\alpha = \left[ (N)^{-1} \sum_{i=1}^N \left( \left| \left( \{X\}_1^N - \left( -\overline{\widehat{ES}^\alpha} \right) \right)^- \right|^p \right) \right]^{\frac{1}{p}},$$
$$\widehat{SDR}^\alpha = \widehat{ES}^\alpha + (1-\alpha)^\beta \widehat{SD}^\alpha. \tag{5}$$

In (5), $N$ is the sample size, $\alpha$ is the significance level, and $\mathbf{1}_*$ is the indication function that assumes the value 1 if * is true and assumes the value 0 if * is false. $(X)^- = \max(-X, 0)$. $\widehat{VaR}^\alpha$ is the negative of the empirical quantile of $X$, $\widehat{ES}^\alpha$ is the negative of the mean below this quantile, $\widehat{SD}^\alpha$ is the semi-deviation at power $p$ below the negative of $\widehat{ES}^\alpha$, and $\widehat{SDR}^\alpha$ is the combination of $\widehat{ES}^\alpha$ and $\widehat{SD}^\alpha$.

As a first visualization, Figure 1 shows the left tail of a hypothetical sample $X \sim N(0.1)$ with VaR, ES and SDR values and with the sign adjusted to $\alpha = 0.01$, $p = 2$ and $\beta = 1$. The measures were calculated based on (5) with a sample size of $N = 10^6$. It is clear that SDR defines superior protection to the ES and, consequently, to VaR. To demonstrate the behavior of SDR, in Figures 2 and 3, we show the evolution of VaR, ES, SD and SDR for different significance levels, also calculated in accordance with (5). Figure 2 represents the Gaussian case $X \sim N(0,1)$ without heavy tails, whereas Figure 3 represents the case with $X \sim t_6$, i.e., with Student's $t$ distribution with heavy tails, which may better represent the behavior of financial assets. We also consider here a sample size of $N = 10^6$. We can observe a common evolution factor among VaR, ES and SDR measures due to the direct relation among their concepts. However, the magnitude of the risk indicated by each measure differs and may represent greater safety offered by the SDR relative to the remaining two measures.

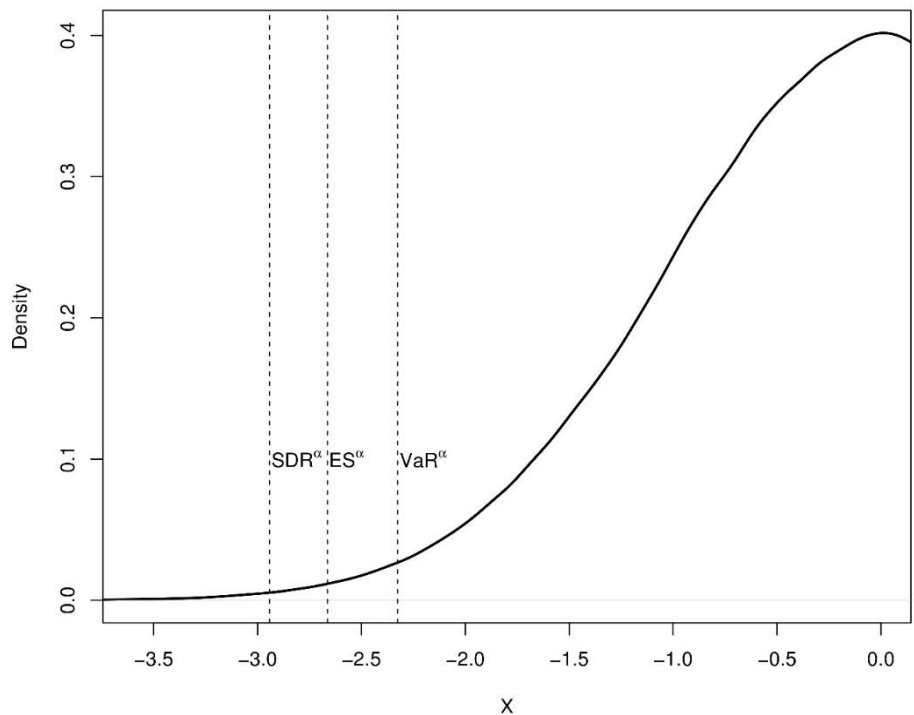

Figure 1 – VaR, ES and SDR, with negative sign, for $\alpha = 0.01$, $p = 2$ and $\beta = 1$ for $X \sim N(0,1)$.

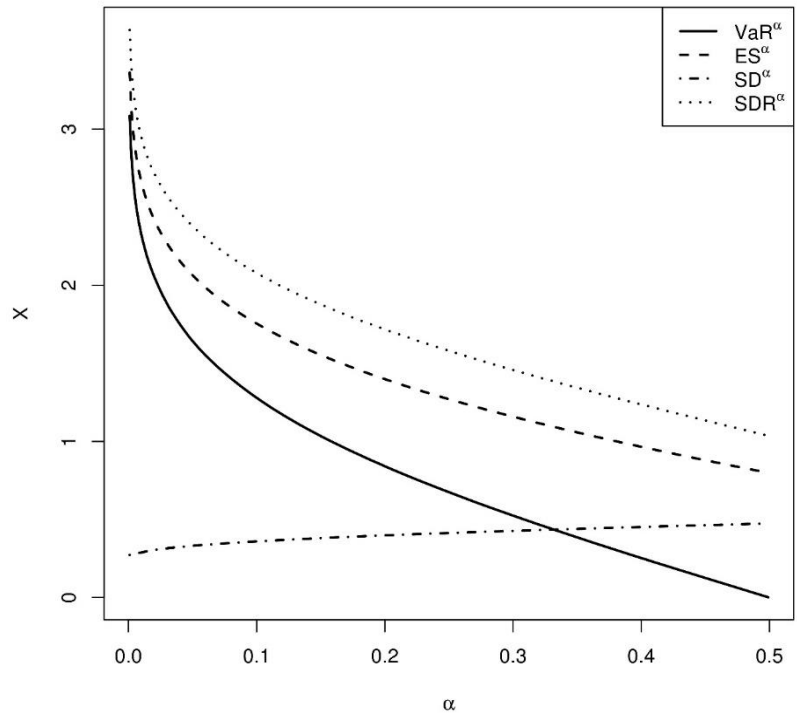

Figure 2 – VaR, ES, SD and SDR as function of $\alpha$ with $X \sim N(0,1)$, $p = 2$ and $\beta = 1$.

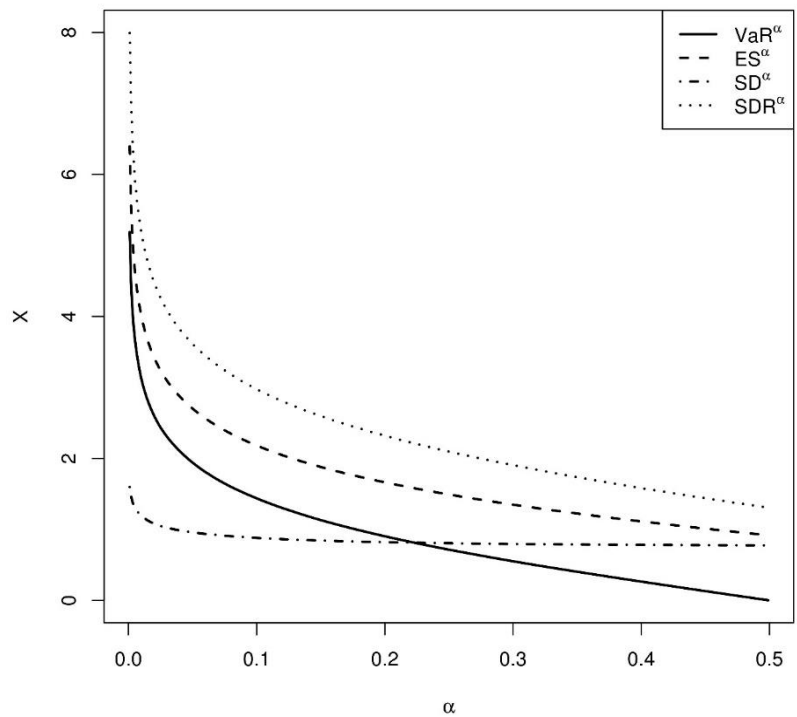

Figure 3 – VaR, ES, SD and SDR as function of $\alpha$ with $X \sim t_6$, $p = 2$ and $\beta = 1$.

The plots contained in Figures 2 and 3 also demonstrate that the measures attain higher values when heavy tails are present. The SDR measure is always above ES, which is above VaR, as previously explained. This difference increases in the direction of most extreme quantiles in the case of Student's t distribution, whereas the opposite behavior is observed for the Gaussian case. This can be explained by the behavior of the tail dispersion, SD, which increases in extreme quantiles in the case of Student's $t$ distribution but decreases in the case of the Gaussian distribution due to the increased probability of occurrence of extreme events between the former and the latter. All measures tend to be equivalent to the values $\sup -X = -\inf X$ when $\alpha$ tends to zero.

Figure 4 shows a three-dimensional plot of the value obtained by the proposed risk measure SDR in relation to values for $0 \leq \alpha \leq 0.50$ and $0 \leq \beta \leq 20$, with $p = 2$, for the case $X \sim t_6$. We also consider here a sample size of $N = 10^6$. The value of the SDR measure increases with lower $\alpha$ and $\beta$ values, which represents more extreme quantiles and greater risk aversion. Figure 4 also reveals an exponential smoothing pattern, which reflects the SD coefficient of penalty on ES.

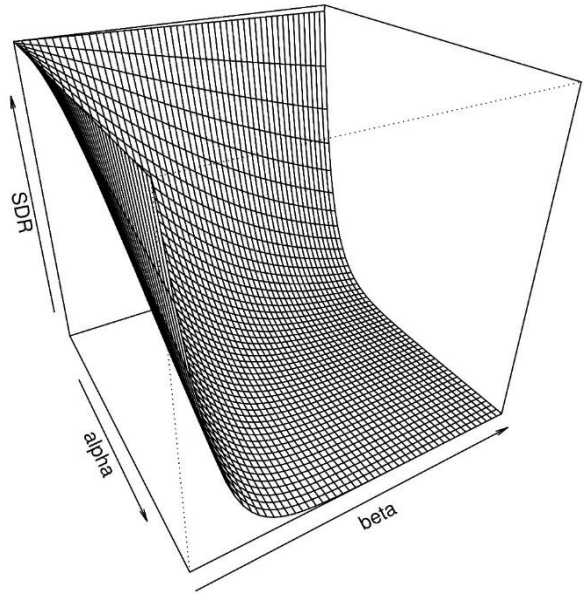

Figure 4 – SDR as a function of $\alpha$ and $\beta$ for $0 \leq \alpha \leq 0,50$, $0 \leq \beta \leq 20$, $p = 2$ and $X \sim t_6$.

To assess the SDR behavior, a more complex analysis is performed by Monte Carlo simulations. The result $X$ is generated by an auto-regressive (AR) process in the conditional mean and generalized auto-regressive conditional heteroscedasticity (GARCH) in the conditional variance AR (1) - GARCH (1,1). This type of specification is frequently considered to analyze risk measures for financial data because it considers stylized facts of daily returns, such as volatility clusters and heavy tails, as noted by Angelidis et al. (2007), among others. The process is parameterized according to (6).

$$X_t = 0.10X_{t-1} + \varepsilon_t, \varepsilon_t = \sigma_t z_t, \ z_t \sim t_v,$$

$$\sigma_t^2 = \sigma^2(1 - 0.10 - 0.85) + 0.10\varepsilon_{t-1}^2 + 0.85\sigma_{t-1}^2. \tag{6}$$

In (6), $X_t$, $\sigma_t^2$, $\varepsilon_t$ and $z_t$ for period $t$ are return, conditional variance, innovation in expectation and a white noise series with Student's $t$ distribution with $E_\mathbb{P}[z_T] = 0$ and $E_\mathbb{P}[(z_T)^2] = 1$, respectively. In addition, $\sigma^2$ is the unconditional (sample) variance. Four scenarios are considered to encompass ($v = 6$) or not encompass ($v = \infty$, i.e., normal distribution) the presence of extreme returns (heavy tails), and periods of low ($\sigma = 0.0125$) and high ($\sigma = 0.022$) volatility. We select the parameters of the data generation process to coincide with the parameters obtained for the daily returns of the North American market index Standard & Poor's 500 (S&P500) before and during the subprime crisis. We attribute this selection to the representativeness of this index, which is also employed in the example with real data and in various simulation studies for risk assessment in finance, such as Christoffersen and Gonçalves (2005) and Degiannakis et al. (2013), among others.

For each scenario (normal distributions and Student's $t$ with low and high volatility), we simulate 10,000 replicates with a sample size of 2,000. This sample size, which represents approximately 8 years of daily observations, is indicated in studies that compare risk measure estimators, such as Kuester et al. (2006), Alexander and Sheedy (2008) and Wong et al. (2012), because it tends to produce lower estimation errors. For each sample, we estimate VaR, ES, SD and SDR using the HS method conform (5). We assume $\beta = 1$ and $p = 2$ to simplify the analysis. All results are presented considering 0.01 and 0.05 as values for $\alpha$ because these values are the most common quantiles in studies and practice. Based on this structure, we calculate the mean values and standard deviation of risk measures estimated with HS for all samples. In addition, we also calculate the mean ratio between each measure and SDR, as well as Pearson's correlation between the values obtained for each measure and SDR. The results of the Monte Carlo simulations are exhibited in Table 1.

Table 1 – Mean, standard deviation, ratio and Pearson's correlation to SDR obtained through Monte Carlo simulation with 10,000 replicates with a sample size of 2,000.

| | Mean | St.Dev. | Ratio | Pearson | | Mean | St.Dev. | Ratio | Pearson |
|---|---|---|---|---|---|---|---|---|---|
| **Normal Distribution, Low Volatility** | | | | | | | | | |
| | | $\alpha = 1\%$ | | | | | $\alpha = 5\%$ | | |
| $\widehat{VaR}^\alpha$ | 0.0352 | 0.0031 | 0.7233 | 0.6932 | $\widehat{VaR}^\alpha$ | 0.0234 | 0.0015 | 0.6181 | 0.6383 |
| $\widehat{ES}^\alpha$ | 0.0431 | 0.0052 | 0.8743 | 0.9054 | $\widehat{ES}^\alpha$ | 0.031 | 0.0029 | 0.8192 | 0.8752 |
| $\widehat{SD}^\alpha$ | 0.0073 | 0.0046 | 0.1262 | 0.8287 | $\widehat{SD}^\alpha$ | 0.007 | 0.0034 | 0.1904 | 0.8883 |
| $\widehat{SDR}^\alpha$ | 0.0494 | 0.0071 | 1.0000 | 1.0000 | $\widehat{SDR}^\alpha$ | 0.038 | 0.0043 | 1.0000 | 1.0000 |
| **Normal Distribution, High Volatility** | | | | | | | | | |
| | | $\alpha = 1\%$ | | | | | $\alpha = 5\%$ | | |
| $\widehat{VaR}^\alpha$ | 0.0624 | 0.0054 | 0.7256 | 0.7025 | $\widehat{VaR}^\alpha$ | 0.0412 | 0.0031 | 0.6191 | 0.6389 |
| $\widehat{ES}^\alpha$ | 0.0766 | 0.0085 | 0.8743 | 0.9092 | $\widehat{ES}^\alpha$ | 0.0543 | 0.0043 | 0.8202 | 0.8787 |
| $\widehat{SD}^\alpha$ | 0.0113 | 0.0068 | 0.1271 | 0.8341 | $\widehat{SD}^\alpha$ | 0.0121 | 0.0042 | 0.1895 | 0.8918 |
| $\widehat{SDR}^\alpha$ | 0.0872 | 0.0122 | 1.0000 | 1.0000 | $\widehat{SDR}^\alpha$ | 0.0675 | 0.0075 | 1.0000 | 1.0000 |
| **Student Distribution, Low Volatility** | | | | | | | | | |
| | | $\alpha = 1\%$ | | | | | $\alpha = 5\%$ | | |
| $\widehat{VaR}^\alpha$ | 0.1631 | 0.8683 | 0.5623 | 0.9831 | $\widehat{VaR}^\alpha$ | 0.0666 | 0.1366 | 0.3986 | 0.9432 |
| $\widehat{ES}^\alpha$ | 0.2453 | 1.3468 | 0.7944 | 0.9952 | $\widehat{ES}^\alpha$ | 0.1277 | 0.5422 | 0.6743 | 0.9978 |
| $\widehat{SD}^\alpha$ | 0.0775 | 0.5587 | 0.2078 | 0.9711 | $\widehat{SD}^\alpha$ | 0.0811 | 0.5344 | 0.3432 | 0.9972 |
| $\widehat{SDR}^\alpha$ | 0.3217 | 1.8752 | 1.0000 | 1.0000 | $\widehat{SDR}^\alpha$ | 0.2044 | 1.0461 | 1.0000 | 1.0000 |

| **Student Distribution, High Volatility** | | | | | | | | | |
|---|---|---|---|---|---|---|---|---|---|
| | | **α = 1%** | | | | | **α = 5%** | | |
| | Mean | St.Dev. | Ratio | Pearson | | Mean | St.Dev. | Ratio | Pearson |
| $\widehat{VaR}^\alpha$ | 0.2792 | 0.5764 | 0.5624 | 0.9776 | $\widehat{VaR}^\alpha$ | 0.1151 | 0.1073 | 0.3977 | 0.9274 |
| $\widehat{ES}^\alpha$ | 0.4191 | 1.0255 | 0.7942 | 0.9932 | $\widehat{ES}^\alpha$ | 0.2192 | 0.3922 | 0.6744 | 0.9883 |
| $\widehat{SD}^\alpha$ | 0.1293 | 0.4596 | 0.2881 | 0.9684 | $\widehat{SD}^\alpha$ | 0.1391 | 0.4711 | 0.3422 | 0.9911 |
| $\widehat{SDR}^\alpha$ | 0.5472 | 1.4582 | 1.0000 | 1.0000 | $\widehat{SDR}^\alpha$ | 0.3513 | 0.8312 | 1.0000 | 1.0000 |

The results in Table 1 indicate different patterns. Note that SDR is more protective than VaR and ES because it exhibits higher mean values. This difference, which is attributed to the SD component, increases in the simulations with Student's $t$ distribution, where more extreme results can occur with higher probability. The SDR is less sensitive to the quantile of interest in relation to the VaR and the ES. Although the risk measured by SDR increases in the 1% quantile, this increase is relatively smaller because SD does not increase in proportion to ES. SD does not always increase in more extreme quantiles. Regarding the deviation, SDR exhibits higher values than the remaining measures, which is natural because it is a combination of the two variables ES and SD and absorbs individual dispersions. The deviation significantly increases for the simulations with Student's $t$ distribution and the 1% quantile. Due to the increased presence of extreme values in these cases, which is the information used to calculate the risk measures, larger oscillations occur, which culminates in a high degree of dispersion.

Regarding the ratio between the measures and the SDR, it is natural to have values smaller than one because the SDR dominates in terms of value obtained compared with the remaining measures. We also verify that the SD component assumes lower values than VaR and ES, and consequently has a lower relative share in the SDR. Concerning the scenarios, due to an increase on SDR in relation to VaR and ES, the ratio increases in the simulations with Student's $t$ distribution, whereas the ratio between SD and SDR increases because SD increases in situations with more extreme values. The SD penalty term becomes very important, especially in scenarios with greater turbulence and a higher probability of large shortfalls. The use of SDR is critical in financial risk management. Regarding correlation, the weakest association is observed between SDR and VaR because ES and SD are direct SDR components. This association becomes significant in scenarios with Student's $t$ distribution. In contexts of higher risk, SDR exhibits higher values compared with ES and VaR, even for measures that capture similar information, and may provide greater protection.

Despite the results obtained with the Monte Carlo simulations, the behavior of SDR when considering real data is important. The use of real data enables the consideration of important occurred events, such as crises and heavy losses. Therefore, we illustrate the application of the SDR in comparison with the most common measures in the risk measurement of the S&P500 index since its creation. As previously mentioned, this indicator is one of the most important indicators for financial markets. Figure 5 presents the temporal evolution of this indicator. By the end of the 1980s, the index experienced mild growth. The trend of sharp rises and sudden drops was caused by financial crises, such as the dot.com and sub-prime crises, in the beginning of the 2000s and the end of the 2000s, respectively. Since then, practices and discussions about risk management, primarily risk measurement, have intensified in the financial area.

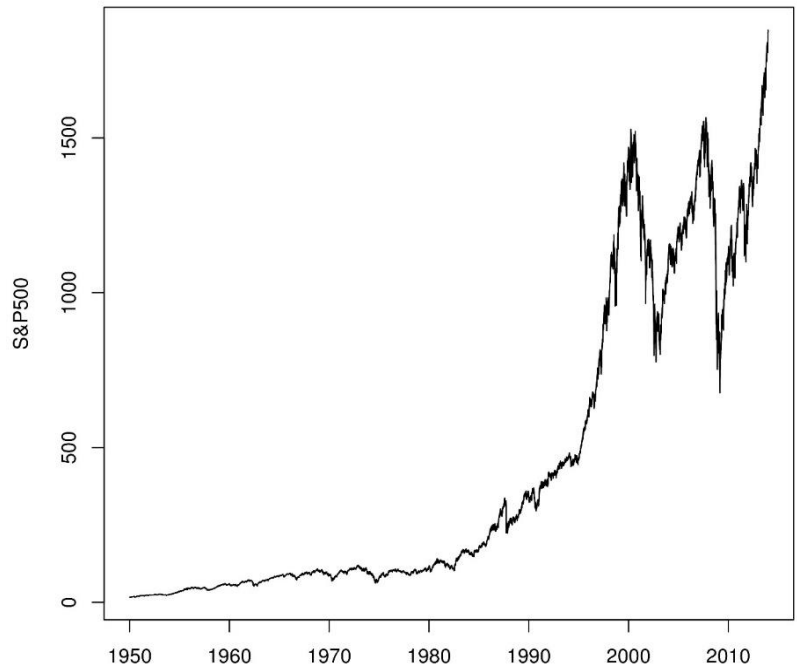

Figure 5 – Daily prices of the S&P500 from January 1950 to December 2013.

To maintain a certain standard for the analyses, the procedure with real data maintains the same structure that we employ for the simulations. We consider the random result $X_t$ as the difference of the natural logarithms or the log difference of the prices in Figure 5. We utilize the HS method to estimate the risk measures according to (5) based on an estimation window of 2,000 observations. To calculate the measures for each day, the last 2,000 observations are employed, where $\beta = 1$, $p = 2$ and 0.01 and 0.05 are considered as values for $\alpha$. Initially, we present a visual analysis of the results. Figures 6 and 7 expose the temporal evolution of S&P500 log-returns, VaR, ES and SDR, with the corrected sign, for the 1% and 5% quantiles. Note that the pattern is very similar between the two quantiles, with a different scale of the measures. The difference between SDR and ES increases at times of major losses and turbulence. By considering the SD dispersion component, SDR has a more solid risk estimate and greater protection during the most critical moments. This behavior is an evident benefit of SDR as a risk measure.

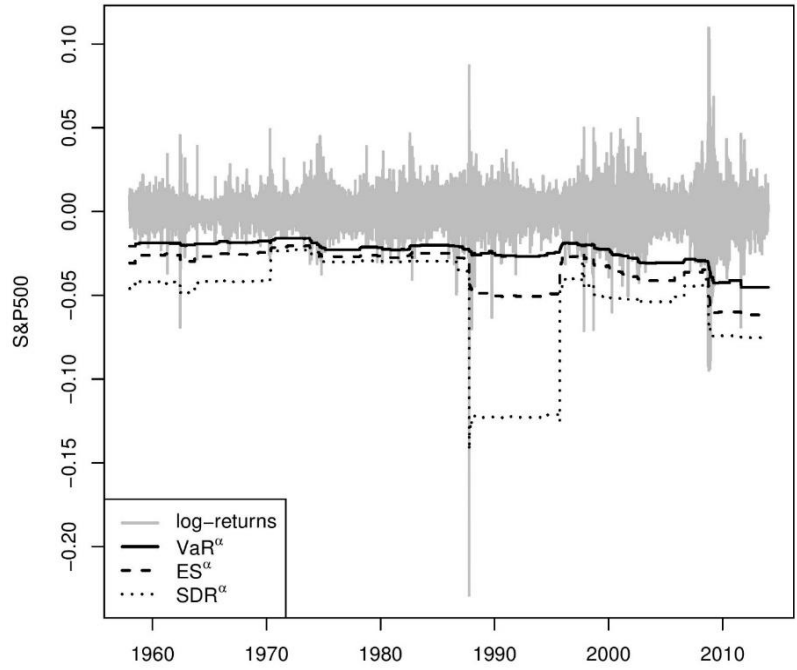

Figure 6 – Daily log-returns of the S&P500, VaR, ES and SDR, with corrected sign, from January 1958 to December 2013 for $\alpha = 0.01$.

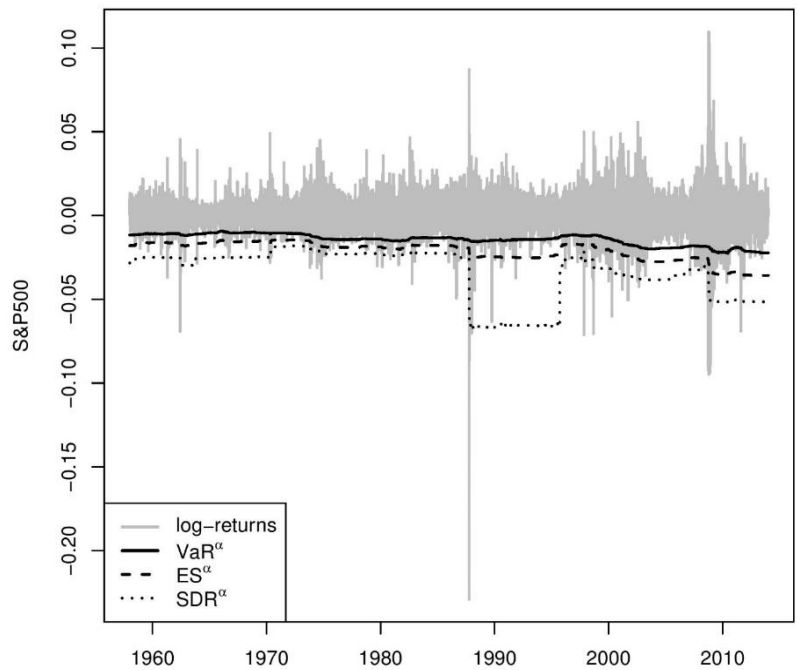

Figure 7 – Daily log-returns of the S&P500, VaR, ES and SDR, with corrected sign, from January 1958 to December 2013 for $\alpha = 0.05$.

In Table 2, we present descriptive statistics of the mean, standard deviation, skewness, kurtosis, minimum and maximum log-returns for the S&P500 and the risk measures estimated with the HS. We also present the mean ratio and Pearson's correlation for SDR. The results in Table 2 confirm that SDR has a higher mean value than the other risk measures because it is more protective. The mean value of the S&P500 log-returns is zero. The difference between SDR and ES is greater than the difference between ES and VaR, implying that the SD penalty represents greater protection. With the exception of SD, the risk measures exhibit higher values for the 1% quantile because it represents more extreme losses. Therefore, SDR does not increase in the same proportion as VaR and ES in the most extreme quantile, which is reflected by the increase in the ratio of the measures in relation to SDR for the 1% quantile. The mean share of SD in the SDR composition decreases for the 1% quantile, but a substantial proportion is retained, which indicates that it cannot be ignored.

Table 2 – Mean, standard deviation, skewness, kurtosis, minumim and maximum, ratio and Pearson's correlation to SDR of the risk measures for the S&P500 from January 1958 to December 2013.

| $\alpha = 1\%$ | Mean | St. Dev. | Skew. | Kurt. | Min. | Max. | Ratio | Pearson |
|---|---|---|---|---|---|---|---|---|
| S&P500 | 0.0002 | 0.0100 | -1.0349 | 30.9024 | -0.2289 | 0.1096 | 0.0056 | 0.0096 |
| $\widehat{VaR}^{\alpha}$ | 0.0242 | 0.0072 | 1.5371 | 4.8403 | 0.0158 | 0.0451 | 0.5283 | 0.4407 |
| $\widehat{ES}^{\alpha}$ | 0.0346 | 0.0121 | 0.9497 | 2.5967 | 0.0203 | 0.0616 | 0.7137 | 0.7987 |
| $\widehat{SD}^{\alpha}$ | 0.0195 | 0.0227 | 1.8176 | 4.7516 | 0.0013 | 0.1027 | 0.2892 | 0.9461 |
| $\widehat{SDR}^{\alpha}$ | 0.0539 | 0.0309 | 1.3893 | 3.6528 | 0.0227 | 0.1426 | 1.0000 | 1.0000 |
| $\alpha = 5\%$ | Mean | St. Dev. | Skew. | Kurt. | Min. | Max. | Ratio | Pearson |
| S&P500 | 0.0002 | 0.0100 | -1.0349 | 30.9024 | -0.2289 | 0.1096 | 0.0008 | 0.0092 |
| $\widehat{VaR}^{\alpha}$ | 0.0143 | 0.0034 | 0.7498 | 2.6015 | 0.0093 | 0.0224 | 0.4632 | 0.4940 |
| $\widehat{ES}^{\alpha}$ | 0.0212 | 0.0058 | 1.0308 | 3.2346 | 0.0143 | 0.0356 | 0.6699 | 0.7144 |
| $\widehat{SD}^{\alpha}$ | 0.0196 | 0.0227 | 1.8176 | 4.7516 | 0.0013 | 0.1027 | 0.4709 | 0.8681 |
| $\widehat{SDR}^{\alpha}$ | 0.0343 | 0.0154 | 1.1129 | 2.7992 | 0.0188 | 0.0674 | 1.0000 | 1.0000 |

With regard to the dispersion of the estimates, the values increase for the 1% quantile, as observed in the results of the Monte Carlo simulations. The range (difference between maximum values and minimum values) maintains this pattern. The kurtosis values also increase in the 1% quantile, with the exception of ES, which exhibits a smoother increase. The observed kurtosis values are similar to the values expected for mesokurtic data, with some deviations. The relationships among dispersion, range and kurtosis are natural because the concepts are correlated. Regarding S&P500 log-returns, the dispersions of the measures are large but the range and kurtosis are significantly smaller, which is natural because returns can assume any value within the empirical data probability distribution, whereas the measures only consider the extreme quantiles.

Regarding the skewness observed in the data, the log-returns have a negative value, which is a stylized fact to financial data, whereas the risk measures have positive value. As the value of the measures have adjusted signs, the skewness values for S&P500 and the risk measures both indicate that major losses occur more frequently than major gains. Regarding the correlations, no linear association between the S&P500 and SDR is evident because the former considers all possible variations and the latter only considers the variations in the tail. The association with VaR is moderate and slightly decreases in the 1% quantile, whereas the association with ES and SD is significant because these two measures compose the SDR and slightly increase in the 1% quantile.

The results obtained with real data are very similar to the results obtained with the Monte Carlo simulation. Due to the considerable length and heterogeneity of the sample, we observe a smoothing effect for extreme losses and volatility, which reveals aspects of the various simulated scenarios in the results for the actual data. Thus, the illustration of SDR use show that the measure confers greater protection than an ES, especially in times or scenarios of greater turbulence when it is needed most. This advantage is acquired because SDR considers the SD dispersion term and considers the two dimensions of the risk concept. In addition to consistent theoretical properties, SDR also has a practical effect on risk measurement.

## 5. Conclusion

Our central focus in this study is to propose the SDR risk measure, which considers the degree of dispersion of an extreme loss in addition to its expected value. SDR is a combination of ES with the SD concept presented in this study and can be defined as the expected loss when this loss exceeds VaR, penalized by the dispersion above this expectation. Therefore, SDR combines two fundamental risk concepts, the probability of bad results (ES) and the variability of an expected result (SD), and considers the tails, which represent extreme results. Thus, SDR has a solid concept and confers more solid protection due to a dispersion penalty.

We discuss definitions and theoretical properties of SDR in detail. Because ES is a known measure, we first demonstrate SD properties. SD is a generalized deviation measure that satisfies Translation Insensitivity, Positive Homogeneity, Subadditivity, Non-Negativity, Lower Range Dominance and Law Invariance axioms. Based on these axioms, we deduce and present the risk envelope and dual representation of SD. Based on SD properties, we obtain theoretical results for SDR. SDR is determined to be a coherent risk measure that satisfies the Translation Invariance, Subadditivity, Monotonicity, Positive Homogeneity, Relevance, Strict Shortfall and Law Invariance axioms. The SDR yields higher values than ES but is limited by the maximum loss and increases in more extreme quantiles, as it is desired for a tail risk measure. Based on these theoretical results, we obtain the dual representation of SDR. We also discuss other issues about SDR, such as the representation via a weighted ES in different quantiles, acceptance sets, convexity and continuity as a function, and relationship with stochastic dominance orders.

We provide illustrations for a better understanding of the concepts, as well as to explore certain practical features of financial risk measurement with the SDR. Thus, we expose plots that help to visualize the SDR in relation to the most common risk measures VaR and ES and the role of the selection of coefficients. Using Monte Carlo simulation procedures and real data, we explore the SDR behavior for different financial scenarios and periods. These results conclude that SDR offers greater protection in risk measurement compared with VaR and ES, especially in times of significant turbulence in riskier scenarios. In these situations, suitable risk management is highly necessary

Based on the theoretical development and examples, SDR is determined to be a risk measure with a solid conceptual foundation, theoretical properties that ensure its use, and high efficiency compared with the most common measures, especially in times of greatest need. In practical terms, the greater usefulness for a risk measure is its application in real problems. We conclude this study with a brief description of the possibilities for the application of the SDR, which can serve as a guide for future studies.

The most evident and direct application is practical risk measurement. The use of SDR has potential in this area because this measure considers the two main pillars of the risk concept, namely uncertainty and the probability of extreme negative results, is coherent, and satisfies the Law Invariance axiom, which enables the measure to be calculated using real data.

Thus, studies that discuss the practical implications of the theoretical properties of the SDR, its role in identifying different types of risk, or even its consistency in the risk management of institutions in relation to other competing measures are examples of possible applications in the field of practical risk measurement.

Another application is the use of risk measures in capital requirement from an institution or agent. This application is closely related to the acceptance set concept, which represents the amount of funds that an institution must maintain to achieve an acceptable position or avoid default. Because SD represents the dispersion around the expected value of the position when an extreme result occurs, greater protection can be achieved by considering the dispersion on the ES as a correction factor, which results in a lower chance of default.

Another field for SDR measure application is resource allocation, which is based on techniques for construction and analyses of portfolios. A fundamental aspect of portfolio optimization is that the objective function must posses the property of convexity. Based on the Positive Homogeneity and Subadditivity axioms, SDR has this property. Therefore, studies that propose minimizing the risk of a portfolio with SDR as an objective function, or even as a restriction for other types of strategies, can contribute to the literature by indicating alternatives for investment analysis based on other risk measures. Similarly, a promising field for application is the decision-making of agents. Due to the continuity properties and the Law Invariance axiom, which are associated with the convexity property, SDR respects the risk aversion of agents. Thus, it is possible to use SDR in the development of models for decision-making.

Other possible applications of SDR in finance are the substitution of other measures in diverse problems. Thus, studies that apply SDR for the development of asset pricing models, the determination of premiums for options or other derivatives, and financial stress diagnosis in turbulent times are recommended.

**Aknowledgements**

We would like to thank an anonymous referee and the editor for their comments. Moreover, we are grateful to the financial support of the CAPES (Brazilian Research Commission) and CNPq (Brazilian Research Council) grant number 552132/2011-0.

**References**

Acerbi, C., 2002. Spectral measures of risk: a coherent representation of subjective risk aversion. Journal of Banking & Finance 26, 7, 1505-1518. doi:10.1016/S0378-4266(02)00281-9

Acerbi, C., Tasche, D., 2002a. On the coherence of expected shortfall. Journal of Banking & Finance 26, 7, 1487-1503. doi:10.1016/S0378-4266(02)00283-2

Acerbi, C., Tasche, D., 2002b. Expected Shortfall: a natural coherent alternative to value at risk. Economic Notes 31, 2, 379-388. DOI: 10.1111/1468-0300.00091

Ahmadi-Javid, A., 2012. Entropic value-at-risk: a new coherent risk measure. Journal of Optimization Theory and Applications 155, 3, 1105-1123. DOI 10.1007/s10957-011-9968-2

Alexander, C., Sheedy, E., 2008. Developing a stress-testing framework based on market risk models. Journal of Banking & Finance 32, 10, 2220-2236. doi:10.1016/j.jbankfin.2007.12.041

Alexander, G. J., Baptista, A. M., 2004. A comparison of VaR and CVaR constraints on portfolio selection with the mean-variance model. Management Science 50, 9, 261-1273. doi: 10.1287/mnsc.1040.0201

Angelidis, T., Benos, A., Degiannakis, S., 2007. A robust VaR model under different time periods and weighting schemes. Review of Quantitative Finance and Accounting 28, 2, 187-201. DOI: 10.1007/s11156-006-0010-y

Artzner, P., Delbaen, F., Eber, J.-M., Heath, D., 1999. Coherent measures of risk. Mathematical Finance 9, 3, 203-228. DOI: 10.1111/1467-9965.00068

Bali, T. G., Demirtas, K. O., Levy, H., 2009. Is there an intertemporal relation between downside risk and expected returns? Journal of Financial and Quantitative Analysis 44, 4, 883-909. doi: 10.1017/S0022109009990159

Bamberg, G., Neuhierl, A., 2010. On the non-existence of conditional value-at-risk under heavy tails and short sales. OR Spectrum 32, 1, 49-60. DOI: 10.1007/s00291-008-0138-3

Bäuerle, N., Müller, A., 2006. Stochastic orders and risk measures: consistency and bounds. Insurance: Mathematics and Economics 38, 1, 132-148. DOI: 10.1016/j.insmatheco.2005.08.003

Belles-Sampera, J., Guillén, M., Santolino, M., 2014. Beyond value-at-risk: GlueVaR distortion risk measures. Risk Analysis 34, 1, 121-134. DOI: 10.1111/risa.12080

Bellini, F., Klar, B., Müller, A., Rosazza Gianin, E., 2014. Generalized quantiles as risk measures. Insurance: Mathematics and Economics 54, 41-48. doi: 10.1016/j.insmatheco.2013.10.015

Belzunce, F., Pinar, J. F., Ruiz, J. M., Sordo, M. A., 2012. Comparison of risks based on the expected proportional shortfall. Insurance: Mathematics and Economics 51, 2, 292-302. doi: 10.1016/j.insmatheco.2012.05.003

Bertsimas, D., Lauprete, G. J., Samarov, A., 2004. Shortfall as a risk measure: properties, optimization and applications. Journal of Economic Dynamics and Control 28, 7, 1353-1381. doi: 10.1016/S0165-1889(03)00109-X

Chen, Z., Wang, Y., 2008. Two-sided coherent risk measures and their application in realistic portfolio optimization. Journal of Banking & Finance 32, 12, 2667-2673. DOI: 10.1016/j.jbankfin.2008.07.004

Chen, Z., Yang, L., 2011. Nonlinearly weighted convex risk measure and its application. Journal of Banking & Finance 35, 7, 1777-1793. doi: 10.1016/j.jbankfin.2010.12.004

Cherny, A. S., 2006. Weighted V@R and its properties. Finance and Stochastics 10, 3, 367-393. DOI: 10.1007/s00780-006-0009-1

Cherny, A. S., Grigoriev, P. G., 2007. Dilatation monotone risk measures are law invariant. Finance and Stochastics 11, 2, 291-298. DOI: 10.1007/s00780-007-0034-8

Christoffersen, P., Gonçalves, S., 2005. Estimation risk in financial risk management. Journal of Risk 7, 3, 1-28.

Cont, R., Deguest, R., Scandolo, G., 2010. Robustness and sensitivity analysis of risk measurement procedures. Quantitative Finance 10, 6, 593-606. DOI: 10.1080/14697681003685597

Cossette, H., Mailhot, M., Marceau, É., Mesfioui, M., 2013. Bivariate lower and upper orthant value-at-risk. European Actuarial Journal 3, 2, 321-357. DOI: 10.1007/s13385-013-0079-3

Cousin, A., Di Bernardino, E., 2013. On multivariate extensions of value-at-risk. Journal of Multivariate Analysis 119, 32-46. doi: 10.1016/j.jmva.2013.03.016

Cousin, A., Di Bernardino, E., 2014. On multivariate extensions of conditional-tail-expectation. Insurance: Mathematics and Economics 55, 272-282. doi: 10.1016/j.insmatheco.2014.01.013

Daníelsson, J., Jorgensen, B. N., Samorodnitsky, G., Sarma, M., De Vries, C. G., 2013. Fat tails, VaR and subadditivity. Journal of Econometrics 172, 2, 283-291. doi: 10.1016/j.jeconom.2012.08.011

Degiannakis, S., Floros, C., Dent, P., 2013. Forecasting value-at-risk and expected shortfall using fractionally integrated models of conditional volatility: international evidence. International Review of Financial Analysis 27, 21-33. doi: 10.1016/j.irfa.2012.06.001

Delbaen, F., 2002. Coherent risk measures on general probability spaces. Advances in Finance and Stochastics, 1-37. DOI: 10.1007/978-3-662-04790-3_1

Dhaene, J., Laeven, R. J. A., Vanduffel, S., Darkiewicz, G., Goovaerts, M. J., 2008. Can acoherent risk measure be too subadditive? The Journal of Risk and Insurance 75, 2, 365-386. DOI: 10.1111/j.1539-6975.2008.00264.x

Dowd, K., Blake, D., 2006. After VaR: the theory, estimation, and insurance applications of quantile-based risk measures. The Journal of Risk and Insurance 73, 2, 193-229. DOI: 10.1111/j.1539-6975.2006.00171.x

Duffie, D., Pan, J., 1997. An overview of value at risk. The Journal of Derivatives 4, 3, 7-49. DOI: 10.3905/jod.1997.407971

El Karoui, N., Ravanelli, C., 2009. Cash subadditive risk measures and interest rate ambiguity. Mathematical Finance 19, 4, 561-590. DOI: 10.1111/j.1467-9965.2009.00380.x

Fischer, T., 2003. Risk capital allocation by coherent risk measures based on one-sided moments. Insurance: Mathematics and Economics 32, 135-146. DOI: 10.1016/S0167-6687(02)00209-3

Föllmer, H., Knispel, T., 2011. Entropic risk measures: coherence vs. convexity, model ambiguity and robust large deviations. Stochastics and Dynamics 11, 02-03, 333-351. Doi: 10.1142/S0219493711003334

Föllmer, H., Schied, A., 2002. Convex measures of risk and trading constraints. Finance and Stochastics 6, 4, 429-447. DOI: 10.1007/s007800200072

Föllmer, H., Schied, A., 2011. Stochastic finance: An introduction in discrete time. 3. ed. Berlin: Walter de Gruyter. DOI: 10.1515/9783110198065

Frittelli, M.; Gianin, E. R., 2002. Putting order in risk measures. Journal of Banking & Finance 26, 1473-1486. doi: 10.1016/S0378-4266(02)00270-4

Furman, E., Landsman, Z. 2006b. On some risk-adjusted tail-based premium calculation principles. Journal of Actuarial Practice 13, 175-190.

Furman, E., Landsman, Z., 2006a. Tail variance premium with applications for elliptical portfolio of risks. ASTIN Bulletin 36, 2, 433-462. DOI: 10.2143/AST.36.2.2017929

Grechuk, B., Molyboha, A., Zabarankin, M., 2009. Maximum entropy principle with general deviation measures. Mathematics of Operations Research 34, 2, 445-467. DOI: 10.1287/moor.1090.0377

Guégan, D., Tarrant, W., 2012. On the necessity of five risk measures. Annals of Finance 8, 4, 533-552. DOI: 10.1007/s10436-012-0205-2

Hamel, A., Rudloff, B., Yankova, M., 2013. Set-valued average value at risk and its computation. Mathematics and Financial Economics 7, 229-246. DOI: 10.1007/s11579-011-0047-0

Inoue, A. K., 2003. On the worst conditional expectation. Journal of Mathematical Analysis and Applications 286, 1, 237-247. doi: 10.1016/S0022-247X(03)00477-3

Jadhav, D., Ramanathan, T., Naik-Nimbalkar, U., 2013. Modified expected shortfall: a new robust coherent risk measure. Journal of Risk 16, 1, 69-83.

Jarrow, R., 2002. Put option premiums and coherent risk measures. Mathematical Finance 12, 2, 135-142. DOI: 10.1111/1467-9965.02003

Jorion, P., 2007. Value at risk: The new benchmark for managing financial risk. 3. ed. Hardcover. DOI: 10.1007/s11408-007-0057-3

Jouini, E., Schachermayer, W., Touzi, N., 2006. Law invariant risk measures have the Fatou property. Advances in Mathematical Economics 9, 49-71. DOI: 10.1007/4-431-34342-3_4

Kaina, M., Rüschendorf, L., 2009. On convex risk measures on $L^p$-spaces. Mathematical Methods of Operations Research 69, 3, 475-495. DOI: 10.1007/s00186-008-0248-3

Kou, S., Peng, X., Heyde, C., 2013. External risk measures and basel accords. Mathematics of Operations Research 38, 3, 393-417. DOI: 10.1287/moor.1120.0577

Krätschmer, V., 2005. Robust representation of convex risk measures by probability measures. Finance and Stochastics 9, 4, 597-608. DOI: 10.1007/s00780-005-0160-0

Krokhmal, P. A., 2007. Higher moment coherent risk measures. Quantitative Finance 7, 4, 373-387. DOI: 10.1080/14697680701458307

Kuester, K., Mittnik, S., Paolella, M., 2006. Value-at-risk prediction: a comparison of alternative strategies. Journal of Financial Econometrics 4, 1, 53-89. doi: 10.1093/jjfinec/nbj002

Kusuoka, S., 2001. On law invariant coherent risk measures. Advances in Mathematical Economics 3, 83-95. DOI: 10.1007/978-4-431-67891-5_4

Lee, J., Prékopa, A., 2013. Properties and calculation of multivariate risk measures: MVaR and MCVaR. Annals of Operations Research 211, 1, 225-254. DOI: 10.1007/s10479-013-1482-5

Leitner, J., 2004. Balayage monotonous risk measures. International Journal of Theoretical and Applied Finance 7, 7, 887-900. DOI: 10.1142/S0219024904002724

Leitner, J., 2005. A short note on second-order stochastic dominance preserving coherent risk measures. Mathematical Finance 15, 4, 649-651. DOI: 10.1111/j.1467-9965.2005.00255.x

Longin, F. M., 2001. Beyond the VaR. The Journal of Derivatives 8, 4, 36-48. DOI: 10.3905/jod.2001.319161

Markowitz, H., 1952. Portfolio selection. The Journal of Finance 7, 1, 77-91. DOI: 10.1111/j.1540-6261.1952.tb01525.x

Pérignon, C., Smith, D. R., 2010. The level and quality of value-at-risk disclosure by commercial banks. Journal of Banking & Finance 34, 2, 362-377. doi: 10.1016/j.jbankfin.2009.08.009

Pflug, G. C., 2000. Some remarks on the value-at-risk and the conditional value-at-risk. Probabilistic Constrained Optimization 49, 272-281. DOI: 10.1007/978-1-4757-3150-7_15

Prékopa, A., 2012. Multivariate value at risk and related topics. Annals of Operations Research 193, 1, 49-69. DOI: 10.1007/s10479-010-0790-2

Pritsker, M., 2006. The hidden dangers of historical simulation. Journal of Banking & Finance 30, 2, 561-582. doi: 10.1016/j.jbankfin.2005.04.013

Righi, M. B., Ceretta, P. S., 2013. Individual and flexible expected shortfall backtesting. Journal of Risk Model Validation 7, 3, 3–20.

Righi, M. B., Ceretta, P. S., 2015. A comparison of Expected Shortfall estimation models. Journal of Economics and Business 78, 14–47. doi: 10.1016/j.jeconbus.2014.11.002

Rockafellar, R. T., Uryasev, S., 2002. Conditional value-at-risk for general loss distributions. Journal of Banking & Finance 26, 7, 1443-1471. doi: 10.1016/S0378-4266(02)00271-6

Rockafellar, R. T., Uryasev, S., Zabarankin, M., 2006. Generalized deviations in risk analysis. Finance and Stochastics 10, 1, 51-74. DOI: 10.1007/s00780-005-0165-8

Sordo, M. A., 2009. Comparing tail variabilities of risks by means of the excess wealth order. Insurance: Mathematics and Economics 45, 3,466-469. doi: 10.1016/j.insmatheco.2009.10.001

Staum, J., 2013. Excess invariance and shortfall risk measures. Operations Research Letters 41, 1, 47-53. doi:10.1016/j.orl.2012.11.004

Svindland, G., 2010. Continuity properties of law-invariant (quasi-)convex risk functions on $L^\infty$. Mathematics and Financial Economics 3, 1, 39-43. DOI: 10.1007/s11579-010-0026-x

Tasche, D., 2002. Expected shortfall and beyond. Journal of Banking & Finance 26, 7, 1519-1533. doi: 10.1016/S0378-4266(02)00272-8

Valdez, E. A., 2005. Tail conditional variance for elliptically contoured distributions. Belgian Actuarial Bulletin 5, 1, 26-36.

Wang, S., 1998. An actuarial index of the right-tail risk. North American Actuarial Journal 2, 2, 88-101. DOI: 10.1080/10920277.1998.10595708

Wong, W. K., Fan, G., Zeng, Y., 2012. Capturing tail risks beyond VaR. Review of Pacific Basin Financial Markets and Policies 15, 03. DOI: 10.1142/S0219091512500154

Wu, G., Xiao, Z., 2002. An analysis of risk measures. Journal of Risk 4, 4, 53-76.

Wylie, J. J., Zhang, Q., Kuen Siu, T., 2010. Can expected shortfall and value-at-risk be used to statically hedge options? Quantitative Finance 10, 6, 575-583. DOI: 10.1080/14697680902956695

Yamai, Y., Yoshiba, T., 2005. Value-at-risk versus expected shortfall: a practical perspective. Journal of Banking & Finance 29, 997-1015. DOI: 10.1016/j.jbankfin.2004.08.010